\documentclass[aps,prb,twocolumn,showpacs,floatfix]{revtex4}
\usepackage{graphicx}
\usepackage{dcolumn}% Align table columns on decimal point
\usepackage{bm}
\usepackage{amsmath}
\newcommand{\nn}{\nonumber}
\newcommand{\ACX}{$A$Cr$_2X_4$}
\newcommand{\ACO}{$A$Cr$_2$O$_4$}
\newcommand{\ACS}{$A$Cr$_2$S$_4$}
\newcommand{\ACSe}{$A$Cr$_2$Se$_4$}
\newcommand{\ACSS}{$A$Cr$_2$S(e)$_4$}
\newcommand{\ZCO}{ZnCr$_2$O$_4$}
\newcommand{\CCO}{CdCr$_2$O$_4$}
\newcommand{\HCO}{HgCr$_2$O$_4$}
\newcommand{\ZCS}{ZnCr$_2$S$_4$}
\newcommand{\CCS}{CdCr$_2$S$_4$}
\newcommand{\HCS}{HgCr$_2$S$_4$}
\newcommand{\ZCSe}{ZnCr$_2$Se$_4$}
\newcommand{\CCSe}{CdCr$_2$Se$_4$}
\newcommand{\HCSe}{HgCr$_2$Se$_4$}
\newcommand{\dxy}{\ensuremath{d_{xy}}}
\newcommand{\dxz}{\ensuremath{d_{zx}}}
\newcommand{\dyz}{\ensuremath{d_{yz}}}
\newcommand{\dzz}{\ensuremath{d_{3z^2-1}}}
\newcommand{\dxxyy}{\ensuremath{d_{x^2-y^2}}}
\newcommand{\tg}{\ensuremath{t_{2g}}}
\newcommand{\tgu}{\ensuremath{t_{2g\uparrow}}}
\newcommand{\tgd}{\ensuremath{t_{2g\downarrow}}}
\newcommand{\eg}{\ensuremath{e_{g}}}
\newcommand{\ag}{\ensuremath{a_{1g}}}

\newcommand{\agd}{\ensuremath{a_{1g\downarrow}}}
\newcommand{\egp}{\ensuremath{e^{\pi}_{g}}}
\newcommand{\egpu}{\ensuremath{e^{\pi}_{g\uparrow}}}

\newcommand{\egs}{\ensuremath{e^{\sigma}_{g}}}
\newcommand{\egsu}{\ensuremath{e^{\sigma}_{g\uparrow}}}
\newcommand{\egsd}{\ensuremath{e^{\sigma}_{g\downarrow}}}
\newcommand{\py}{\ensuremath{p_{y}}}
\newcommand{\px}{\ensuremath{p_{x}}}
\newcommand{\pz}{\ensuremath{p_{z}}}

\newcommand{\tdds}{\ensuremath{t_{dd\sigma}}}
\newcommand{\tddp}{\ensuremath{t_{dd\pi}}}
\newcommand{\tddd}{\ensuremath{t_{dd\delta}}}
\newcommand{\ep}{\ensuremath{\varepsilon_{p}}}
\newcommand{\ed}{\ensuremath{\varepsilon_{d}}}
\newcommand{\dcc}{\ensuremath{d_{\text{Cr--Cr}}}}
\newcommand{\dcx}{\ensuremath{d_{\text{Cr--}X}}}
\newcommand{\dax}{\ensuremath{d_{A\text{--}X}}}
\newcommand{\kv}{\ensuremath{\mathbf{k}}}
\newcommand{\mb}{\ensuremath{\mu_{\text{B}}}}
\newcommand{\kb}{\ensuremath{k_{\text{B}}}}
\newcommand{\ef}{\ensuremath{E_{F}}}
\newcommand{\LDAU}{LSDA+$U$}
\newcommand{\mv}{\ensuremath{\mathbf{m}}}
\newcommand{\qv}{\ensuremath{\mathbf{q}}}
\newcommand{\rv}{\ensuremath{\mathbf{r}}}
\newcommand{\tv}{\ensuremath{\mathbf{t}}}

\newcommand{\Rv}{\ensuremath{\mathbf{R}}}
\newcommand{\Sv}{\ensuremath{\mathbf{S}}}
\newcommand{\Ebq}{\ensuremath{E_{bnd}(\mathbf{q})}}
\newcommand{\Etq}{\ensuremath{E_{tot}(\mathbf{q})}}
\newcommand{\Eq}{\ensuremath{E(\mathbf{q})}}
\newcommand{\EHq}{\ensuremath{E_{\text{H}}(\mathbf{q})}}
\newcommand{\TCW}{\ensuremath{\Theta_{\text{CW}}}}
\newcommand{\eau}{\ensuremath{\varepsilon^{ab}_{\uparrow}}}

\newcommand{\ebu}{\ensuremath{\varepsilon^{b}_{\uparrow}}}

\newcommand{\dtud}{\ensuremath{\Delta_{\uparrow\downarrow}}}
\newcommand{\deud}{\ensuremath{\Delta'_{\uparrow\downarrow}}}
\newcommand{\dcf}{\ensuremath{\Delta_{CF}}}

\newcommand{\degt}{\ensuremath{\Delta_{e-t}}}
\newcommand{\JH}{\ensuremath{J_{\text{H}}}}
\newcommand{\sigv}{\ensuremath{\bm{\sigma}}}

\begin{document}

\title{Electronic band structure and exchange coupling constants in $A$Cr$_2X_4$ spinels}

\author{A.~N. Yaresko}

\affiliation{Max Planck Institute for the Physics of Complex
Systems, D-01187 Dresden, Germany}

\date{\today}

\begin{abstract}
We present the results of band structure calculations for \ACX\ ($A$=Zn,
Cd, Hg and $X$=O, S, Se) spinels. Effective exchange coupling constants
between Cr spins are determined by fitting the energy of spin spirals to a
classical Heisenberg model. The calculations reproduce the change of the
sign of the dominant nearest-neighbor exchange interaction $J_1$ from
antiferromagnetic in oxides to ferromagnetic in sulfides and selenides. It
is verified that the ferromagnetic contribution to $J_1$ is due to indirect
hopping between Cr \tg\ and \eg\ states via $X$ $p$ states.
Antiferromagnetic coupling between 3-rd Cr neighbors is found to
be important in all the \ACX\ spinels studied, whereas other interactions
are much weaker. The results are compared to predictions based on the
Goodenough-Kanamori rules of superexchange.
\end{abstract}


\maketitle

\section{\label{sec:introd}Introduction}

Chromium spinels provide unrivalled possibilities for studying magnetic
interactions in solids. In these compounds with a general formula \ACX,
where $A$ is a divalent nonmagnetic cation (Mg, Zn, Cd, or Hg) and $X$ is a
divalent anion (O, S, or Se), a Cr$^{3+}$ ion is in the 3$d^3$
configuration. Its three 3$d$ electrons occupy the majority-spin states of
a completely spin polarised \tg\ sub-shell leading to the total spin
$S$=3/2. Although charge and orbital degrees of freedom in the \ACX\
spinels are frozen, these compounds show wide variety of magnetic
properties ranging from those of a strongly frustrated antiferromagnet to a
Heisenberg ferromagnet.  Depending on the chemical composition their
effective Curie-Weiss temperature (\TCW) varies from $-$400 K in oxides to
200 K in selenides,\cite{BWRL66,RKMH+07} which indicates that the sign of
the dominant exchange interaction changes from antiferromagnetic (AFM) to a
ferromagnetic (FM) one.

In the \ACO\ spinels AFM nearest-neighbor interactions between Cr spins
residing on a pyrochlore lattice are geometrically frustrated. The magnetic
ground state of a frustrated antiferromagnet is highly degenerate which
leads to unusual low-temperature properties. \cite{Ramirez94,MC98} Cr oxide
spinels remain paramagnetic with the Curie-Weiss form of the magnetic
susceptibility down to temperatures well below $|\TCW|$ of 398, 71, and 32
K for $A$=Zn, Cd, and Hg. \ZCO\ ($T_{N}$=12.5 K) and \CCO\ ($T_{N}$=7.8 K)
undergo a first order phase transition of the Spin-Peierls type into a
magnetically ordered N\'eel state at temperatures much lower than the
characteristic strength $|\TCW|$ of the interaction between Cr
spins.\cite{LBKR+00,UKMG+05} The transitions are accompanied by cubic to
tetragonal structural distortions, however, the sign of the distortions and
the magnetic order below $T_{N}$ are different. In \ZCO\ the lattice
contracts along the $c$ axis ($c<a$) and the N\'eel state has a complex
commensurate spin structure with 4 characteristic wave vectors. In
contrast, the lattice of \CCO\ expands below $T_{N}$ ($c>a$) and its
ordered state is incommensurate with a wave vector
$\mathbf{Q}$=(0,$\delta$,1) with $\delta\sim$0.09.\cite{CMLK+05} \HCO\ also
undergoes a transition to a magnetically ordered state at $T_{N}$=5.8 K but
the symmetry of the lattice lowers to orthorhombic. \cite{UMGU06}
Recently, a metamagnetic transition and a wide magnetisation plateau with
the magnetic moment equal to one-half of the full Cr moment have been
observed in \CCO\ and \HCO. \cite{UKMG+05,UMGU06,PSS04,UIMG+07}

In \ACSS\ spinels dominant ferromagnetic interactions are not geometrically
frustrated. Nevertheless, Cr spins in \ZCS\ and \ZCSe\ form helical spin
structures below 15.5 and 18 K, respectively. \cite{HHSC95,ASSI+78,HRKM+06}
In \ZCS\ the helical structure coexists at low temperatures with a
collinear AFM one. The transitions into helically ordered state are
supposed to occur because of competing FM nearest-neighbor interaction and
AFM interactions between more distant Cr neighbors. Recently, it has been
shown that \CCS\ and \HCS\ exhibit ferroelectric behavior with strong
increase of the dielectric constant below the temperature of magnetic
ordering. \cite{HLFK+05,WLFH+06}

The diversity of magnetic properties of the \ACX\ spinels can hardly be
explained without understanding the mechanism of exchange interactions
between Cr spins, their range, and relative strengths. So far theoretical
analyses of the effective exchange interactions in Cr spinels were mostly
based on the Goodenough-Kanamori rules of
superexchange. \cite{Goodenough58,Kanamori59} J.~Goodenough in Ref.\
\onlinecite{Goodenough69} explained the FM sign of the nearest-neighbor
coupling $J_1$ in \ACSS\ by indirect hopping between half-filled Cr \tg\
and empty \eg\ states via $p$ states of $X$ anions.  K.~Dwight and
N.~Menyuk (Refs.\ \onlinecite{DM67,DM68}) analyzed various superexchange
paths for interactions between up to 6-th Cr neighbors and concluded that
AFM coupling constants $J_3$ between 3-rd neighbors and even weaker FM
$J_4$ and $J_6$ may be relevant alongside $J_1$. Then, the estimated $J_n$
were used to examine the stability of different spiral ground states in
\ZCSe. To our knowledge the only attempt to obtain the values of $J_n$ from
\textit{ab initio} band structure calculations was made in Ref.\
\onlinecite{CFT06} where the coupling constants $J_1$--$J_3$ were
calculated by comparing the total energies of several simple spin
configurations.

The aim of the present work is to compare electronic band structures of
\ACX\ ($A$=Zn, Cd, Hg and $X$=O, S, Se) spinels calculated within the local
spin density (LSDA) as well as \LDAU\ approximations and to estimate
exchange coupling constants between Cr spins by fitting the calculated
energy of spin spirals to a classical Heisenberg model. The paper is
organized as follows. The spinel crystal structure is shortly described in
Sec.\ \ref{sec:structure}. Some details of the calculational procedure are
given in Sec.\ \ref{sec:details}. In Sec.\ \ref{sec:bandstr} the band
structures of the \ACX\ spinels calculated using LSDA and \LDAU\ are
compared and their dependence on the chemical composition is analyzed. The
results on the exchange coupling constants $J_n$ are presented in Sec.\
\ref{sec:exch}. The comparison of calculated $J_n$ to experimental data and
the discussion of their origins are given in Sec.\
\ref{sec:discus}. Finally, the results are summarised in Sec.\
\ref{sec:summ}.

\section{\label{sec:structure}Crystal structure}

\begin{figure}[tbp!]
\begin{center}
\includegraphics[width=0.44\textwidth]{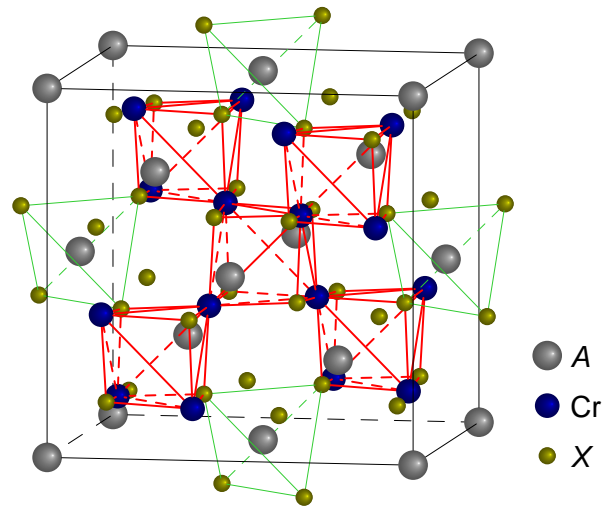}
\end{center}
\caption{\label{fig:ACrX_str}(Color online) The crystal structure of \ACX\
spinels with an $A$ ion shifted to the origin. Distorted Cr$_4X_4$ cubes
and $AX_4$ tetrahedra are plotted by thick (red) and thin (green) lines,
respectively.}
\end{figure}

\ACX\ compounds considered here belong to a large family of
$A^{2+}B^{3+}_2X^{2-}_4$ spinels which crystallise to a cubic $Fd\bar{3}m$
(N227) structure; with $A$, $B$, and $X$ ions occupying $8a$ (1/8,1/8,1/8),
$16d$ (1/2,1/2,1/2), and $32e$ ($x$,$x$,$x$) Wyckoff positions,
respectively. The experimental values of the lattice constant ($a_0$) and
fractional coordinates ($x$) are collected in Table~\ref{tab:str}.  The
spinel crystal structure plotted in Fig.\ \ref{fig:ACrX_str} can be
considered as built of distorted Cr$_4X_4$ cubes which share a Cr site. Cr
and $X$ ions that belong to the same cube form two regular tetrahedra with
a common center, which coincides with the center of the cube. Each cube is
linked via $X$ ions to four regular $AX_4$ tetrahedra. The centers of the
Cr$_4X_4$ cubes and $AX_4$ tetrahedra form two diamond lattices shifted by
a vector (1/4,1/4,1/4). Finally, Cr ions are arranged along chains running
in $\langle110\rangle$ directions and form the so-called pyrochlore lattice
which consists of corner-sharing regular tetrahedra.

As the ionic radius of a $A^{2+}$ cation increases in the row
Zn$^{2+}\rightarrow$Cd$^{2+}\rightarrow$Hg$^{2+}$, $X$ ions are pushed
further away from the center of a $AX_4$ tetrahedron. This leads to lattice
expansion and corresponding increase of Cr--Cr distances. The distortion
of Cr$_4X_4$ cubes also increases, which allows to avoid too strong
elongation of Cr--$X$ bonds and manifests itself in the increase of the $x$
parameter.  When O$^{2-}$ ions are replaced by S$^{2-}$ or Se$^{2-}$, which
have significantly larger ionic radii, the length of both $A$--$X$ and
Cr--$X$ bonds increases. This results in even stronger increase of the
lattice constant. The nearest Cr--Cr (\dcc), Cr--$X$ (\dcx), and $A$--$X$
(\dax) distances are summarized in Table~\ref{tab:str}.

\begin{table}
\caption{\label{tab:str} Experimental lattice constants $a_0$ (\AA) and
fractional coordinate $x$ of $X$ ions used in the calculations and the
shortest Cr--Cr, Cr--$X$, and $A$--$X$ distances (\AA). The last column
contains experimental values of \TCW\ taken from Ref.\
\onlinecite{RKMH+07}.}
\begin{ruledtabular}
\begin{tabular}{cD{.}{.}{2.3}D{.}{.}{1.4}D{.}{.}{1.3}D{.}{.}{1.3}D{.}{.}{1.3}r}
 Compound &  \multicolumn{1}{c}{$a$} & \multicolumn{1}{c}{$x$}
& \multicolumn{1}{c}{\dcc} & \multicolumn{1}{c}{\dcx} 
& \multicolumn{1}{c}{\dax} & \TCW\\
\hline
ZnCr$_2$O$_4$\footnotemark[1]  & 8.327 & 0.2616 & 2.944 & 1.990 & 1.970 &-398\\
CdCr$_2$O$_4$\footnotemark[2]  & 8.600 & 0.2682 & 3.041 & 2.006 & 2.133 & -71\\
HgCr$_2$O$_4$\footnotemark[2]  & 8.661 & 0.2706 & 3.062 & 2.003 & 2.184 & -32\\
ZnCr$_2$S$_4$\footnotemark[3]  & 9.982 & 0.2619 & 3.529 & 2.383 & 2.367 & 7.9\\
CdCr$_2$S$_4$\footnotemark[4]  &10.240 & 0.2647 & 3.620 & 2.419 & 2.478 & 90\\
HgCr$_2$S$_4$\footnotemark[5]  &10.256 & 0.267  & 3.626 & 2.402 & 2.523 & 140\\
ZnCr$_2$Se$_4$\footnotemark[6] &10.484 & 0.2599 & 3.707 & 2.522 & 2.450 & 155\\
CdCr$_2$Se$_4$\footnotemark[7] &10.735 & 0.2642 & 3.795 & 2.540 & 2.588 & 184\\
HgCr$_2$Se$_4$\footnotemark[5] &10.737 & 0.264  & 3.796 & 2.543 & 2.585 & 200\\
\end{tabular}
\end{ruledtabular}
\footnotetext[1]{H.~Sawada, Ref.~\onlinecite{Sawada97}.} 
\footnotetext[2]{H.~Ueda, Ref.~\onlinecite{UIMG+07}.}
\footnotetext[3]{E.~Riedel and E.~Horvath, Ref.~\onlinecite{RH69}.}
\footnotetext[4]{T.~Borovskaya \textit{et al.}, Ref.~\onlinecite{BBTP+91}.}
\footnotetext[5]{J.~Hemberger \textit{et al.}, Ref.~\onlinecite{RKMH+07}.}
\footnotetext[6]{J.~Akimitsu \textit{et al.}, Ref.~\onlinecite{ASSI+78}.}
\footnotetext[7]{J.~Krok-Kowalski \textit{et al.}, Ref.~\onlinecite{KRGW+95}.}
\end{table}

Each Cr site is surrounded by a trigonally distorted $X_6$ octahedron; with
all Cr--$X$ distances in the octahedron being equal. The degree of the
trigonal distortion is determined by the value of the fractional coordinate
$x$. For $x>$0.25 the octahedron is expanded along one of the
$\langle111\rangle$ directions and becomes regular for $x$=1/4.

The local symmetry of a Cr site is $D_{3d}$. It is worth noting that since
six Cr ions nearest to a Cr site form a trigonal antiprism, the symmetry
remains trigonal even when $x$ is equal to the ideal value of 1/4.

\section{\label{sec:details}Computational details}

The calculations of the electronic band structure of the \ACX\ spinels were
performed for the experimentally observed lattice constant and $X$
fractional coordinates (Table \ref{tab:str}) using the linear muffin-tin
orbital (LMTO) method \cite{And75} with the combined correction terms taken
into account. In order to decrease overlap between atomic spheres 3 sets of
empty spheres ($E$) were added at $8b$ (0,0,0), $16c$ (3/8,3/8,3/8), and
$48f$ ($x'$,1/8,1/8) Wyckoff positions. Muffin-tin orbitals with angular
momentum $l\le2$ for $A$, Cr, and $X$ spheres and $l\le1$ for empty spheres
were included into the basis set. The Perdew--Wang parameterisation
\cite{PW92} for the exchange-correlation potential in the local
spin-density approximation was used. Brillouin zone (BZ) integrations were
performed using the improved tetrahedron method. \cite{BJA94}

Calculations for spiral spin structures based on the generalized Bloch
theorem \cite{San91a} were performed under the assumption that the
direction of the magnetisation \mv\ is constant within each atomic
sphere. Inside a sphere at $\tv+\Rv$, where \tv\ defines its position in a
unit cell and \Rv\ is a lattice vector, the magnetisation direction is
determined by an angle $\theta_t$ between the spin moment and $z$ axis, the
wave vector \qv\ of a spiral, and a phase $\phi_t$:
\begin{equation}
\mv(\rv_t)=m(r_t)\left(\begin{array}{c}
\cos (\phi_t + \qv\cdot\Rv) \sin \theta_t \\
\sin (\phi_t + \qv\cdot\Rv) \sin \theta_t \\
\cos \theta_t \\
\end{array}\right),
\end{equation}
with $\rv_t=\rv-\tv-\Rv$.

When performing LMTO calculations for spin spirals it is convenient to
split the LSDA exchange-correlation potential into a spin- and
\qv-independent part $V=(V_{\uparrow}+V_{\downarrow})/2$ and an effective
exchange field $B=(V_{\uparrow}-V_{\downarrow})/2$, where $V_{\uparrow}$
and $V_{\downarrow}$ are exchange-correlation potentials for majority- and
minority-spin electrons defined in the site-dependent local spin frame, in
which the spin-density matrix for a given site is diagonal. In the present
calculations \emph{spin-independent} LMTO basis functions were constructed
starting from the solution of the Kohn-Sham equation with only the
spin-independent part $V$ of the exchange-correlation potential included to
the LSDA one electron potential. Matrix elements of the spin-dependent
part of the LMTO Hamiltonian
\begin{eqnarray}
&&H_{B}=\sum_{\tv,\Rv} \mathbf{B}(\rv_t) \cdot \sigv 
=\sum_{\tv}B(r_t)\times \nn \\  
&&\left(
\begin{array}{cc}
\cos \theta_t  & \sin \theta_t \sum_{\Rv} e^{-i(\phi_t+\qv\cdot\Rv)} \\
\sin \theta_t \sum_{\Rv} e^{i(\phi_t+\qv\cdot\Rv)} & -\cos \theta_t 
\end{array}
\right),
\label{eq:hb}
\end{eqnarray}
where \sigv\ is a vector of Pauli matrices, were included at a variational
step. Radial matrix elements of $B(r)$ between the solution of the
Kohn-Sham equation inside a sphere $\phi_{\nu}(r)$ and/or its energy
derivative $\dot{\phi}_{\nu}(r)$ were calculated by numerical integration.

The diagonal in spin indices matrix elements of $H_B$ calculated between
two Bloch wave functions with wave vectors \kv\ and $\kv'$ do not depend on
\qv\ and, as usual, are non-zero only if $\kv=\kv'$. In the absence of the
spin-orbit coupling the only off-diagonal in spin indices terms of the
Hamiltonian are $H^{\downarrow\uparrow}_B(\qv)$ and
$H^{\uparrow\downarrow}_B(\qv)$ given by Eq.\ (\ref{eq:hb}), which couple
the states with $\kv'-\kv=\pm\qv$. Then, the LMTO Hamiltonian matrix can be
written in the following block form:
\begin{equation}
H = \left(\begin{array}{cc}
H^{\downarrow\downarrow}_{\kv-\qv/2,\kv-\qv/2} & 
H^{\downarrow\uparrow}_{\kv-\qv/2 , \kv+\qv/2} \\
H^{\uparrow\downarrow}_{\kv+\qv/2 , \kv-\qv/2} & 
H^{\uparrow\uparrow}_{\kv+\qv/2 , \kv+\qv/2}
\end{array}\right).
\label{eq:hkkq}
\end{equation}
Thus, for a spin spiral with an arbitrary \qv\ one needs to diagonalise one
$2N\times2N$ matrix instead of two $N\times N$ matrices, one for each spin
channel, in conventional spin-polarised calculations. \cite{San91a} If,
however, the spin-orbit coupling term is included into the Hamiltonian it
additionally couples majority- and minority-spin states with the same $\kv$
and the Hamiltonian matrix becomes infinite.

Most calculations reported in Sec.\ \ref{sec:exch} were performed for
planar spin spirals with all $\theta_t=\pi/2$. Then, for a given \qv\ the
magnetisation direction inside a sphere is defined solely by the phase
$\phi_t$. The phases at Cr sites were fixed by requiring that
$\phi_t=\qv\cdot\tv$. At other atomic and empty sites they were determined
self-consistently. At each iteration of a self-consistency loop a rotation
to the local spin frame in which the spin-density matrix becomes diagonal
was found and the corresponding rotation angles were used to determine new
magnetisation direction for the next iteration. \cite{BK98} Iterations were
repeated until self-consistency in the electron spin density as well as in
the magnetisation direction in each atomic sphere was achieved.

Finally, in order to account for correlation effects in the Cr 3$d$ shell
we adopted the \LDAU\ method \cite{AZA91} in the rotationally invariant
representation. \cite{LAZ95,YAET+00} The so-called atomic limit \cite{CS94}
was used for the double counting term. The effective screened Coulomb
repulsion $U$ between 3$d$ electrons was considered as a parameter of the
model and varied from 2 to 4 eV. For the on-site exchange integral \JH\ the
value of 0.9 eV estimated from the LSDA calculations was used.

\section{\label{sec:bandstr}Comparison of the electronic structures of O-, S-,
and S\lowercase{e}-based spinels}

\subsection{\label{sec:lda}Spin-restricted LDA results and hopping matrix
elements} 

\begin{figure}[tbp!]
\begin{center}
\includegraphics[width=0.45\textwidth]{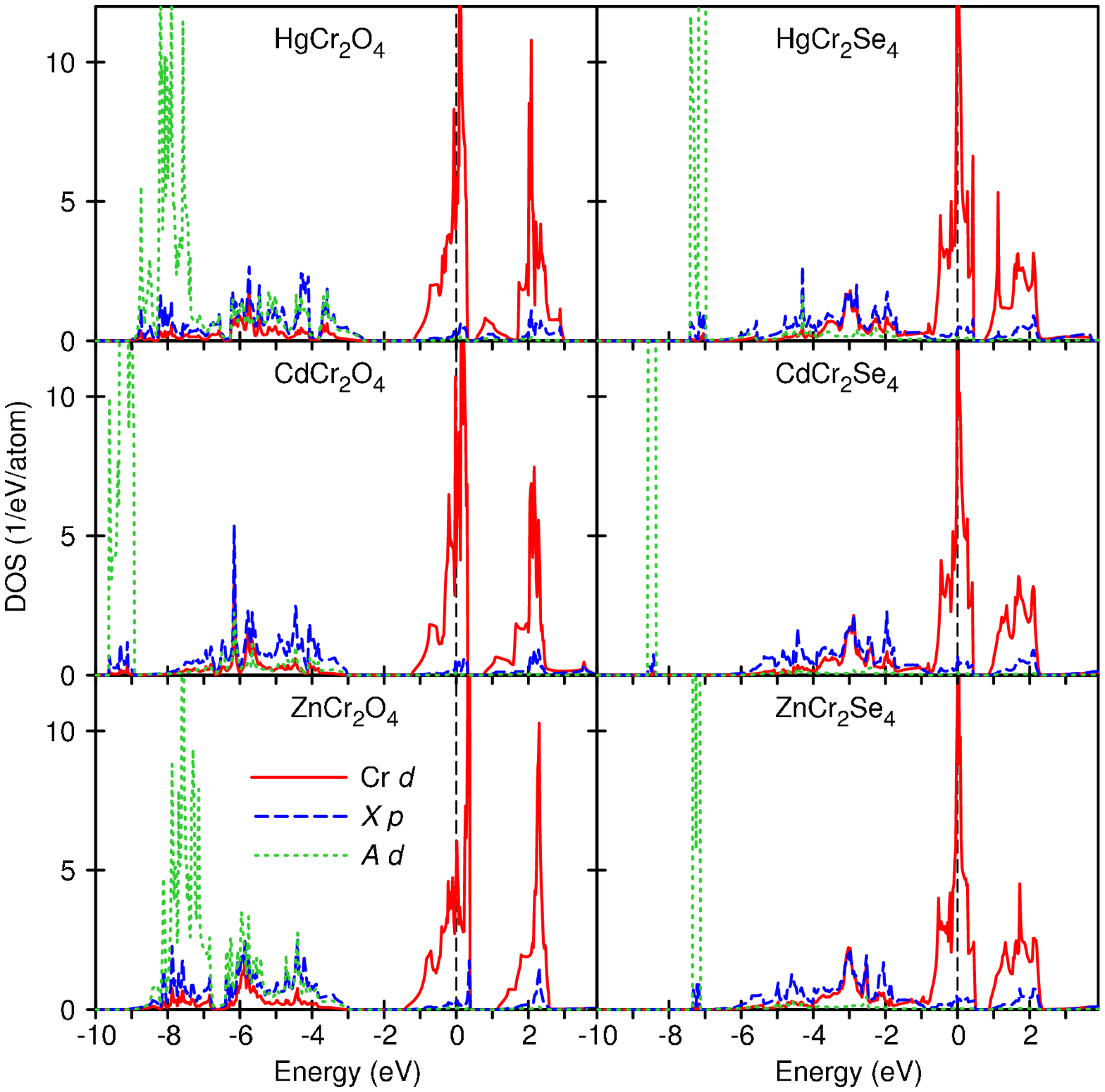}
\end{center}
\caption{\label{fig:ACrX_dos}(Color online) Densities of Cr $d$ (red solid
lines), $X$ $p$ (blue dashed lines), and $A$ $d$ (green dotted lines)
states in $A$Cr$_2$O$_4$ and $A$Cr$_2$Se$_4$. Zero energy is chosen at the
Fermi level.}
\end{figure}

The effect of the chemical composition on the electronic structure of \ACX\
spinels can be analyzed by comparing the densities of Cr $d$ and $X$ $p$
states obtained from spin-restricted LDA calculations for $A$Cr$_2X_4$
(Fig.\ \ref{fig:ACrX_dos}). Since site-resolved densities of states (DOS)
for the $A$Cr$_2$S$_4$ spinels show the same trends as those for
$A$Cr$_2$Se$_4$, only DOS calculated for $A$Cr$_2$O$_4$ and $A$Cr$_2$Se$_4$
compounds are presented in Fig.\ \ref{fig:ACrX_dos}.

In all the \ACX\ spinels considered in the present work occupied bands in
the energy range down to $-$10 eV below the Fermi level (\ef) originate
from $A$ $d$, Cr $d$, and $X$ $p$ states. $A$ $d$ states in the sulfides
and selenides as well as Cd 4$d$ states in \CCO\ form a narrow group of
bands below the bottom of $X$ $p$ states. Zn $3d$ and Hg $5d$ bands in the
corresponding oxides cross the bottom of O $p$ states and hybridise
strongly with the latter which results in the appearance of wide peaks of
$d$ DOS between $-9$ and $-7$ eV. The $A$ $d$ states are completely
occupied and lie well below \ef\ so that they have no effect on the
magnetic properties of the \ACX\ spinels.

O $p$-derived bands in \ACO\ spread over the energy range from $-$8 to $-3$
eV and are separated by a gap of $\sim1.5$ eV from Cr 3$d$ states which
give prevailing contribution to the bands crossing the Fermi level and a
prominent DOS peak at \ef. Because of weaker electronegativity of S$^{2-}$
and Se$^{2-}$ ions as compared to O$^{2-}$, $X$ $p$ states in \ACSS\
spinels move closer to \ef\ and form bands between $-6.5$ and $-0.5$ eV. In
contrast to oxides the top of $X$ $p$ bands overlaps with the bottom of Cr
$d$ ones and the gap between $X$ $p$ and Cr $d$ states closes.

Cr $d$ states are split by the cubic component of the crystal field at a Cr
site into a triplet \tg\ (\dxy, \dxz, and \dyz) and a doublet \eg\ (\dzz\
and \dxxyy). In a Cr$X_6$ octahedron the \tg\ and \eg\ states form
relatively weak $pd\pi$- and much stronger $pd\sigma$-type bonds with the
$X$ $p$ states, respectively. Bonding Cr $d$ -- $X$ $p$ combinations
participate in the formation of the $X$ $p$-derived bands which is
evidenced by the rather high density of Cr $d$ states in this energy range
(see Fig.\ \ref{fig:ACrX_dos}). These states are completely filled and
stabilise the Cr$X_6$ octahedron. In all the \ACX\ spinels considered here,
the partially occupied bands crossing \ef\ are formed by antibonding
combinations of the Cr $d$ \tg\ and $X$ $p$ states with the dominant
contribution of the former.  These states, filled with 3 electrons, play
the crucial role in the formation of Cr magnetic moments and effective
exchange interactions between them.  The Cr $d$ states of the \eg\ symmetry
which form antibonding $pd\sigma$ combinations with the $X$ $p$ states are
shifted to higher energies and separated by an energy gap from the
antibonding Cr \tg\ -- $X$ $p$ states.  For the sake of brevity in the
following these bands are referred to simply as Cr $d$ \tg\ and \eg\ bands.

\begin{figure}[tbp!]
\begin{center}
\includegraphics[width=0.44\textwidth]{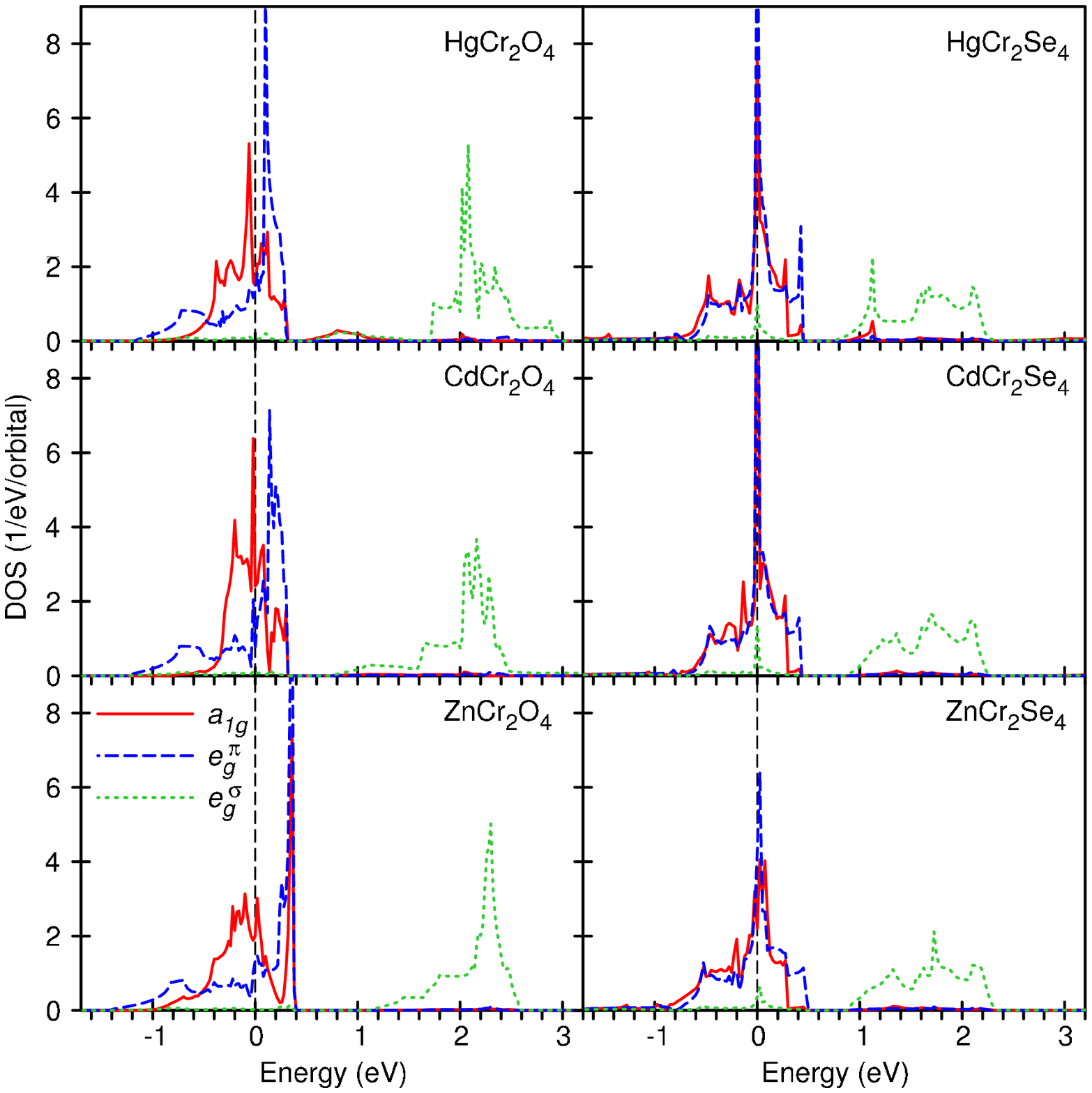}
\end{center}
\caption{\label{fig:Cr_d_dos}(Color online) Symmetry-resolved densities of
the Cr $d$ \ag\ (red solid lines), \egp\ (blue dashed lines), and \egs\
(green dotted lines) states in $A$Cr$_2$O$_4$ and $A$Cr$_2$Se$_4$. Zero
energy is chosen at the Fermi level.}
\end{figure}

As the local symmetry of a Cr site is lower than cubic the \tg\ states are
additionally split into a singlet and a doublet which transform according
to \ag\ and \eg\ representations of the $D_{3d}$ group. We denote this
doublet as \egp\ in order to distinguish it from the doublet formed by
\dzz\ and \dxxyy\ orbitals (\egs) which transforms according to the same
\eg\ representation. The densities of the Cr $d$ states of \ag, \egp, and
\egs\ symmetries are shown in Fig.\ \ref{fig:Cr_d_dos}.

The trigonal splitting between the \ag\ and \egp\ states is much smaller
than their bandwidths; with the center of gravity of the former being about
0.05 eV lower. The corresponding DOS curves in \ACO\ are, however,
remarkably different. The total width of the Cr \tg\ sub-band is determined
by the \egp\ states. Their DOS curve is very asymmetric with low DOS at the
bottom and a huge DOS peak at the top of the \tg\ bands. The \ag\ states
are significantly narrower. They are responsible for the DOS peak at \ef\
but contribute also to the high-energy \egp\ DOS peak. This strong
suppression of the width of the \ag\ DOS is a characteristic feature of
$3d$ transition metal oxides with the spinel structure and a small ionic
radius of an $A$ ion. Recently, two scenarios based on a strong Coulomb
interaction within V $d$ shell and the localized behavior of V \ag\ states
have been proposed to explain heavy-fermion-like properties in
LiV$_2$O$_4$. \cite{NPKP+03,AHLA07}

A simplified analysis using Slater--Koster integrals \cite{SK54} shows that
the narrowing of \ag\ DOS is caused by a delicate balance between direct
$d$--$d$ hopping matrix elements $t_{dd}$ between the Cr \tg\ states and an
effective indirect hopping $t^{p}_{dd}$ via O $p$ states. Under the
assumption that $\tddd=\tdds/6$, $\tddp=-2\tdds/3$, and $\tdds<0$ the
strongest hopping between \tg\ states of nearest Cr ions is a negative
$dd\sigma$-type hopping between those orbitals which have their lobes
directed along one of Cr chains, e.g.\ a \dxy--\dxy\ hopping along the
[110] direction. Hoppings of $dd\pi$ type between the other two \tg\
orbitals (\dyz\ and \dxz) along the same chain are positive and about two
times weaker, whereas \dxy--$d_{yz/zx}$ hoppings are zero. After making
transformation to the symmetrised combinations of \tg\ orbitals it turns
out that the matrix elements between \ag\ and \egp\ orbitals on neighboring
Cr sites are much smaller than the matrix elements between the states of
the same symmetry, with the \ag--\ag\ hopping being stronger than the
\egp--\egp\ one. This results in almost decoupled \ag\ and \egp\ bands,
with the bandwidth of the former being \emph{larger} than of the latter.

The contribution of the hopping via O $p$ states to the lowest order in
$t^2_{pd}/(\ed-\ep)$, where $t_{pd}$ is the Cr $d$--O $p$ matrix element
and \ep\ and \ed\ are the energies of the Cr $d$ and O $p$ states, can be
easily estimated using the L\"owdin partitioning technique. \cite{Low63I}
In the ideal ($x=1/4$) spinel structure with regular Cr$_4$O$_4$ cubes the
indirect hopping contributes only to \dyz--\dxz\ and \dxz--\dyz\
hybridisations via O \pz\ states. The corresponding matrix element
$t^p_{yz,xz}$ is negative, so that the direct and indirect terms compensate
each other. However, as it was shown for LiV$_2$O$_4$ in Ref.\
\onlinecite{EHHL+99} and verified for \ZCO\ in the present work, the
densities of \ag\ and \egp\ states calculated for a hypothetical spinel
structure with $x=1/4$ are of the same width. Thus, the reduction of the
\dyz--\dxz\ hopping is not sufficient to explain the narrowing of the \ag\
states. When the distortion of Cr$_4X_4$ cubes is taken into account,
non-zero indirect hoppings proportional to the degree of the distortion
$\delta=x-1/4$ appear between all pairs of \tg\ orbitals. However, the most
significant one is an indirect \dxy--\dxy\ matrix element between the
orbitals pointing along Cr chains. Already for $x$=0.26 it becomes as large
as 0.6$t^p_{yz,xz}$ and, being of the opposite sign, strongly suppresses
the direct \dxy--\dxy\ hopping. The net effect of the competition between
the direct and indirect contributions to the effective $d$--$d$ hopping is
the reduction of the \ag\ bandwidth relative to the width of \egp\
states. If, however, the direct hopping becomes weak, the width of the \ag\
and \egp\ states is governed by the indirect hopping via $X$ $p$ states and
depends strongly on the deviation of the fractional coordinate $x$ from the
ideal value of 1/4.

Finishing the discussion of the $d$--$d$ hybridisation we have to mention
strong hopping matrix elements between \tg\ and \egs\ states which appear
as a result of either direct hopping between, for instance, \dxy\ and \dzz\
orbitals in the [110] direction or indirect one via \px\ and \py\ states of
two $X$ ions closest to the Cr--Cr bond. The indirect hopping is allowed
already in an undistorted Cr$_4X_4$ cube and is stronger than $t^p_{yz,xz}$
for one of the involved $p$--$d$ bonds is of $pd\sigma$ type. However, the
direct and indirect contributions are of opposite signs which may lead to
strong suppression of effective \tg--\egs\ hopping matrix elements.
Finally, an indirect contribution to hopping between \egs\ states of
nearest Cr ions is proportional to $\delta$ and is of the same sign as the
corresponding direct hopping.

Let us come back to the results of band structure calculations shown in
Figs.\ \ref{fig:ACrX_dos} and \ref{fig:Cr_d_dos}. As Zn is replaced by a
larger Cd or Hg ion, the width of the Cr \tg\ states decreases from 1.8 eV
in \ZCO\ to 1.6 eV in \HCO. The narrowing of the bands can be explained by
the reduction of the strength of both $d$--$d$ and Cr $d$--O $p$ hopping
matrix elements caused by the lattice expansion (see Table
\ref{tab:str}). The shape of \ag\ and \egp\ DOS, however, also varies
considerably. This indicates the change of relative strengths of the direct
$d$--$d$ and indirect $d$--O $p$--$d$ hoppings between Cr \tg\ states.

Cr \egs\ states are affected much stronger by the change of the chemical
composition because of their interaction with $s$ bands of an $A$ ion. In
\ZCO\ \egs\ bands cross just the bottom of a Zn 4$s$ band near the BZ
center. Cd 5$s$ and, especially, Hg 6$s$ bands shift to lower energies and
overlap with Cr \egs\ bands. In \CCO\ the hybridisation with a Cd 5$s$ band
is responsible for a low energy tail of Cr \egs\ DOS. In \HCO\ the bottom
of a Hg 6$s$ band comes so close to Cr \tg\ bands that it starts to
hybridise via O $p$ states with Cr \ag\ bands, which is evidenced by the
appearance of noticeable density of \ag\ and \egs\ states in the energy gap
between \tg\ and \egs-derived bands.

Symmetry resolved densities of Cr $d$ states in selenides, presented on the
right hand side of Fig.\ \ref{fig:Cr_d_dos}, are strikingly different from
the corresponding DOS curves in oxides. The densities of \ag\ and \egp\
states do not differ so much as in oxides; with the top of the \egp\ bands
being shifted to somewhat higher energies. The huge DOS peak found at the
very top of the \tg\ states in \ZCO\ and slightly lower in the two other
oxides shifts in selenides closer to the center of the \tg\ bands. The
energy difference between the Cr \egs\ and \tg\ states is somewhat smaller
than in oxides. The \egs\ DOS curves become wider and more symmetric.
Another important distinction of the \ACSS\ electronic structure is the
enhanced weight of the \egs\ states in the Bloch wave functions of Cr \tg\
bands, which is revealed by noticeable density of the \egs\ states in the
corresponding energy range.

\begin{figure}[tbph!]
\begin{center}
\includegraphics[width=0.44\textwidth]{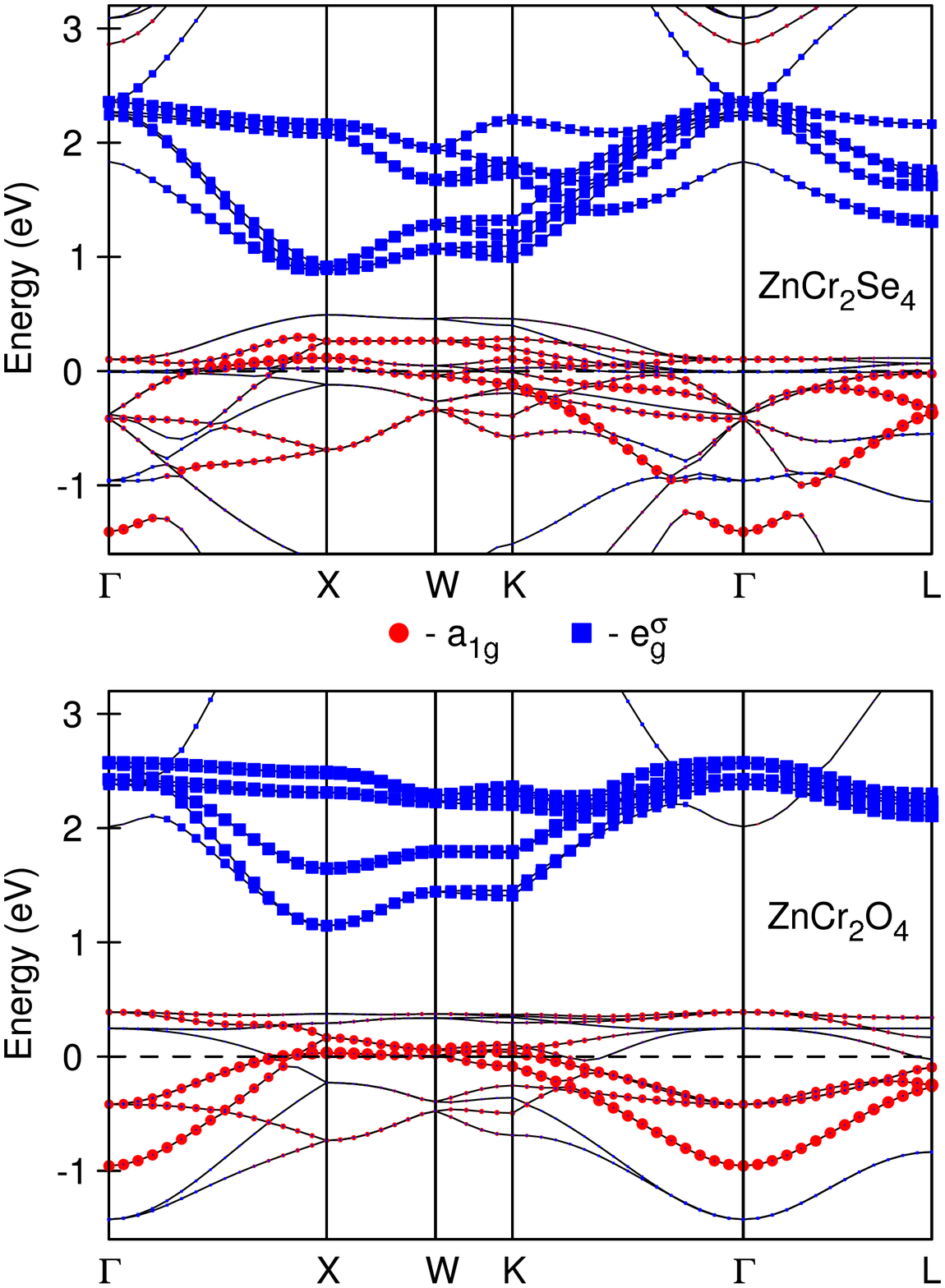}
\end{center}
\caption{\label{fig:FB_Zn}(Color online) ``Fat'' bands calculated along
some symmetry lines in the fcc BZ for \ZCO\ (lower panel) and \ZCSe\ (upper
panel). The size of filled (red) circles and (blue) squares is proportional
to the partial weight of Cr \ag\ and \egs\ states in the Bloch wave
function, respectively. Zero energy is chosen at the Fermi level.}
\end{figure}

The difference between the dispersions of Cr $d$ bands in the oxides and
selenides is illustrated in Fig.\ \ref{fig:FB_Zn} which shows ``fat'' bands
calculated for \ZCO\ and \ZCSe; with the size of filled circles and squares
being proportional to the partial weight of Cr \ag\ and \egs\ states,
respectively, in the Bloch wave function at a given \kv-point.

The abovementioned differences in the band dispersions and DOS suggest that
the effective $d$--$d$ hopping matrix elements should also be significantly
different, which is not surprising taking into account the increase of the
lattice constants in \ACSS\ as compared to \ACO\ (Table \ref{tab:str}). The
increase of Cr--Cr distances leads to strong reduction of the direct
$d$--$d$ hopping matrix elements which fall off as $1/\dcc^5$. It should be
noted, however, that a tight binding analysis of the band structure of
\CCS\ performed in Ref.\ \onlinecite{SMS00} showed that the direct hopping
between Cr $d$ states is not negligible even in sulfides and selenides.
The contribution of the indirect hopping via $X$ $p$ states to the
effective $d$--$d$ hoppings, which is proportional to $t_{pd}^2/(\ed-\ep)$,
does not change much because i) S 3$p$ and Se 4$p$ states are significantly
more spatially extended than O 2$p$ states and, although \dcx\ also
increases in \ACSS, Cr $d$--$X$ $p$ hoppings $t_{pd}$ remain approximately
of the same strength as in \ACO; ii) the energy difference $\ed-\ep$ in
sulfides and selenides is smaller than in oxides.

Similar to oxides the Cr \tg\ states in selenides become narrower with the
increase of the $A$ ionic radius. The hybridisation of Cr \egs\ states with
Hg 6$s$ bands leads to sharpening of a DOS peak near the bottom of the
\egs\ sub-band.

\subsection{\label{sec:lsda}The effect of spin-polarisation}

\begin{figure}[tbp!]
\begin{center}
\includegraphics[width=0.45\textwidth]{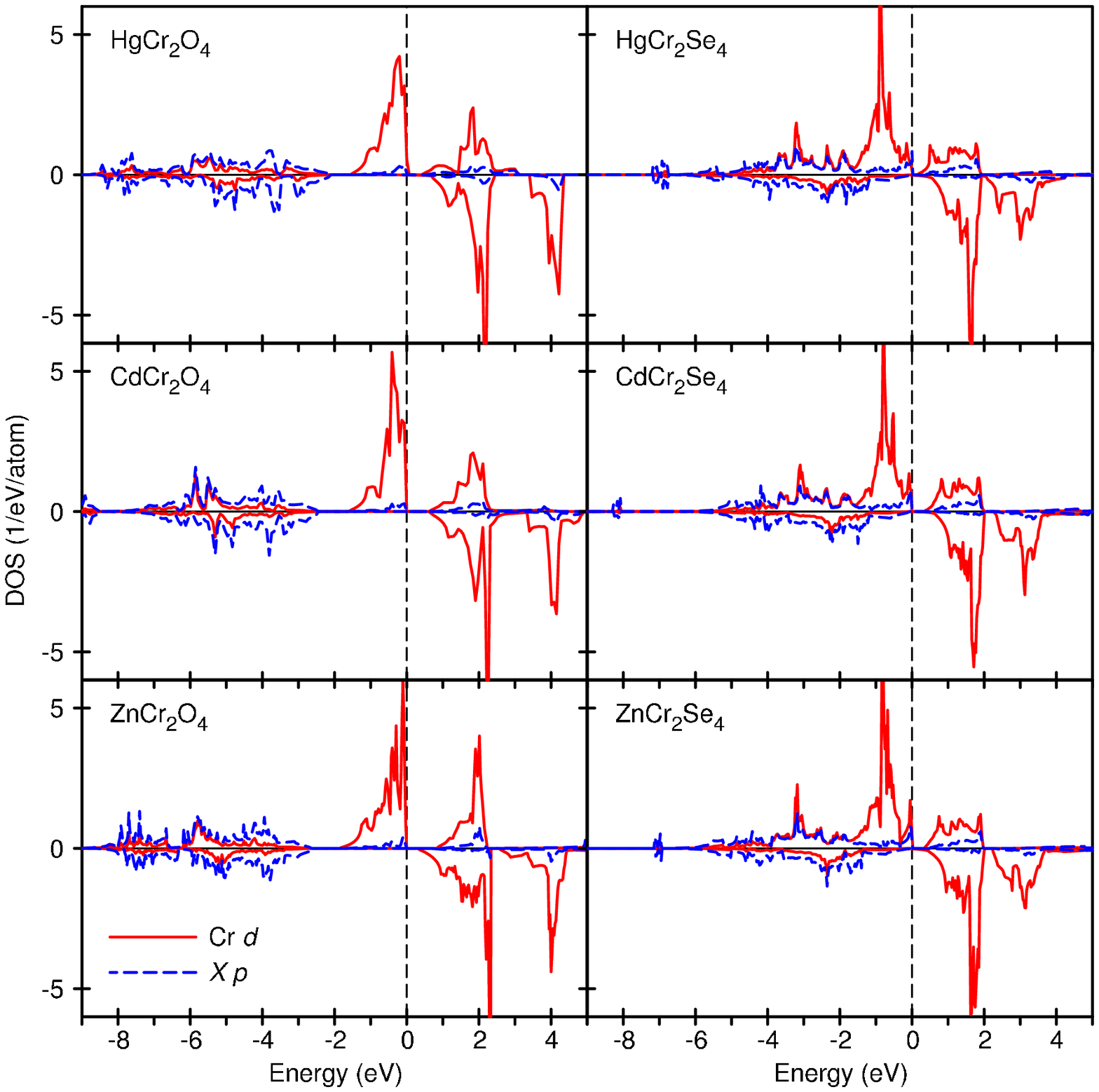}
\end{center}
\caption{\label{fig:ACrX_fm}(Color online) Cr $d$ (red solid lines) and $X$
$p$ (blue dashed lines) DOS in $A$Cr$_2$O$_4$ and $A$Cr$_2$Se$_4$ obtained
from spin-polarized calculations with FM alignment of Cr magnetic
moments. Zero energy is at the Fermi level.}
\end{figure}

When in spin-polarized LSDA calculations Cr $d$ states with two spin
projections are allowed to have different occupations, the strong on-site
exchange interaction splits the half-filled Cr \tg\ shell into occupied
majority-spin \tgu\ and unoccupied minority-spin \tgd. The densities of Cr
$d$ and $X$ $p$ states in \ACO\ and \ACSe\ obtained from spin-polarized
calculations with the ferromagnetic (FM) alignment of Cr moments are shown
in Fig.\ \ref{fig:ACrX_fm}.

In all the spinels studied here the exchange splitting of Cr $d$ states is
about 2.7 eV which gives an estimate of 0.9 eV for the on-site Hund's
exchange coupling \JH. The exchange and crystal field splittings are of
comparable strengths, so that Cr \tgd\ states are found at the same energy
as \egsu. In the oxides \tgu\ states remain separated by an energy gap from
O $p$ bands. In \ZCO\ and \CCO\ the FM solution is insulating with
completely occupied \tgu\ and empty \tgd\ states. In \HCO\ the bottom of a
minority-spin Hg $s$ band hybridised with Cr \agd\ states crosses \tgu\
bands leading to a metallic solution.

In contrast to the oxides, in the \ACSS\ spinels \tgu\ states move below
the top of the $X$ $p$ states.  As a result the highest occupied
majority-spin bands are formed mainly by $X$ $p$ states strongly hybridised
with Cr \egpu. In the Zn and Hg spinels they overlap with the very bottom
of mostly unoccupied \tgd\ bands while for \CCS\ and \CCSe\ insulating
solutions with a tiny gap were obtained. The LSDA band structures and DOS
calculated for \CCS\ and \CCSe\ are in good agreement with the results of
previous calculations. \cite{SMS00,CPMS94} In the present work, however,
the FM solution for \CCS\ is insulating whereas in Ref.\ \onlinecite{SMS00}
metallic FM solutions were obtained for both compounds. A possible cause
for the difference is that in Ref.\ \onlinecite{SMS00} structural data with
the S fractional coordinate $x$=1/4 were used.

A Cr spin magnetic moment defined as a volume integral of the spin density
over a Cr atomic sphere varies from 2.86 \mb\ in \ZCO\ to 3.15 \mb\ in
\CCSe, which agrees with a good accuracy with the value of 3 \mb\ expected
for the completely polarized \tg\ shell filled with 3 electrons. The
hybridisation with Cr $d$ states induces weak negative spin-polarisation of
$X$ $p$ states. It is, however, partially compensated by a positive
contribution coming from the ``tails'' of Cr $d$ states that have the $d$
symmetry inside an $X$ atomic sphere. The net spin magnetic moment $M_X$
induced on $X$ ions is small and negative with $M_{\text{O}}\approx-0.01$
\mb, $M_{\text{S}}\approx-0.05$ \mb, and $M_{\text{Se}}\approx-0.09$
\mb. The moment induced on $A$ ions is less than 0.05 \mb.

The self-consistent FM solution discussed above can be obtained for any of
the 9 spinels but it does not necessarily have the lowest total energy. If
the coupling between Cr spins is antiferromagnetic (AFM) then, because of
geometrical frustrations intrinsic to the pyrochlore lattice formed by Cr
ions, any configuration of classical spins such that the sum of 4 spin
vectors sitting in the corners of each tetrahedron is equal to zero is the
ground state of the Heisenberg Hamiltonian with nearest neighbor
interactions. \cite{MC98} The requirement of zero total spin in each
tetrahedron is satisfied, for example, if Cr spins are aligned
ferromagnetically along the [1$\pm$10] chains and antiferromagnetically
along [01$\pm$1] and [10$\pm$1] ones or, in other words, if Cr spins in
each (001) plane are parallel but the spins in consecutive planes along the
[001] direction are antiparallel to each other. In the following this spin
configuration will be denoted as AFZ. It is easy to check that in this
case only 4 of 6 Cr--Cr bonds in each tetrahedron are AFM while the other
two remain FM and the magnetic energy cannot be minimised for all 6 bonds
simultaneously if the coupling between Cr spins is antiferromagnetic.

Another spin configuration which gives zero total spin in each tetrahedron
and does not break the cubic symmetry of the lattice is a non-collinear one
with Cr spins directed along the lines passing through the center of a
tetrahedron, i.e., along one of the $\langle111\rangle$ directions; with
all four Cr spins pointing either to or away from the
center. Self-consistent band structure calculations performed with this
configuration of Cr spins gave for all the \ACX\ spinels insulating
solutions with zero net magnetic moment, which in the following is referred
to as a ZM solution. Cr magnetic moments calculated in the local spin frame
are about 0.1 \mb\ smaller than for the corresponding FM solution but still
close to 3 \mb.  Depending on the magnetisation directions of four Cr ions
in a Cr$_4X_4$ cube the magnetisations in four $X$ atomic spheres are
parallel or antiparallel to the vectors pointing to the center of the cube
so that small $X$ magnetic moments also cancel each other. With this
arrangement of Cr and $X$ moments the spin moments of $A$ ions are equal to
zero. It is worth noting that for all the \ACX\ compounds the LSDA total
energy difference between the AFZ and ZM solutions does not exceed 3 meV
per formula unit.

The comparison of the LSDA total energies of FM and ZM solutions shows that
in \ZCO\ the solution with zero net moment is significantly more
favourable; with the energy difference per formula unit (f.u.) being 171
meV. In \CCO\ this difference decreases to 19 meV, whereas in \HCO\ the two
solutions are almost degenerate; the FM one being 2 meV lower. In the
sulfides and selenides the FM solution becomes more preferable. As Zn is
replaced by a heavier ion the energy difference between the two solution
increases from 43 to 118 meV/f.u.\ in sulfides and from 80 to 134 meV/f.u.\
in selenides.

These results indicate that the change of the sign of dominant exchange
interactions from AFM in oxides to FM in sulfides and selenides is captured
already by LSDA band structure calculations. Moreover, all the compounds
show a clear tendency to ferromagnetism as the lattice expands due to the
increase of the radius of $A$ ions.

\subsection{\label{sec:lsdau}\LDAU\ results}

In \ACX, as well as in many other 3$d$ compounds, LSDA underestimates
correlation effects in the rather localised Cr 3$d$ shell. The strong
electronic correlations can be accounted for, at least on a mean-field
level, by using the \LDAU\ approach. Since Cr \tg\ states are split by the
on-site exchange interaction into occupied majority- and unoccupied
minority-spin states, charge and orbital degrees of freedom are frozen
already in LSDA. As a consequence, \LDAU\ band structures of the \ACX\
spinels are qualitatively similar to the LSDA results.

When non-spherical terms of the Coulomb repulsion are neglected, the
orbital dependent \LDAU\ potential $V_i$ can be approximated as
$V_i=U'(1/2-n_i)$, where $n_i$ is the occupation of $i$-th localised
orbital, $U'=U-\JH$, and $U$ is the effective screened Coulomb repulsion
between 3$d$ electrons. Taking into account that the orbital occupations
$n_i$ of Cr \tgu\ and \tgd\ states are close to 1 and 0, respectively, the
main effect of the \LDAU\ potential is to shift the \tgu\ states to lower
and \tgd\ to higher energies by $U'/2$.  Thus, the splitting between the
\tgd\ and \tgu\ states increases from 3\JH\ in LSDA calculations to
$3\JH+U'$ in \LDAU\ ones. In oxides, because of the downward shift of the
\tgu\ states, they start to overlap with O $p$ bands. In \ACSS\ the \tgu\
states move further below the top of $X$ $p$ bands and their contribution
to the highest occupied majority spin bands decreases. Due to the increase
of the \tgd--\tgu\ splitting, ferromagnetic \LDAU\ solutions for all 9
compounds considered in the present work become insulating starting from
$U=$3 eV.

Because of the strong Cr \egs--$X$ $p$ hybridisation discussed in Sec.\
\ref{sec:lda} the orbital occupations $n_i$ calculated for the formally
unoccupied Cr \egs\ states are in the range 0.28--0.48 for the majority-
and about 0.22--0.25 for the minority-spin orbitals. As a result the shift
of the Cr \egs\ states to higher energies caused by the Coulomb repulsion
$U$ is somewhat smaller than for the \tgd\ states for which $n_i$ does not
exceed 0.07 and the splitting between \egsd\ and \tgd\ states decreases
with the increase of $U$.

\section{\label{sec:exch}Effective exchange coupling constants}

Effective exchange coupling constants $J_n$ between Cr spins were
determined by performing band structure calculations for spiral spin
structures with a varying wave vector \qv. Then, the \qv-dependence of the
energy of the spin spirals was mapped onto a classical Heisenberg model.
In order to get reliable estimates for $J_n$ one needs to perform
calculations for a sufficiently large number of \qv. Since self-consistent
calculations for spin spirals are much more time-consuming than
conventional collinear spin-polarised calculations, the \qv-dependence of
their energy was calculated using the so-called local force theorem
(LFT). \cite{LKAG87,ST98} According to this approach the total energy
difference between two spin configurations can be approximated by the
difference of their band energies, provided that the calculations are
performed starting from the same electron spin densities and only the
magnetisation direction varies. Strictly speaking this approximation is
justified only for small deviations of the magnetisation direction from
some collinear spin arrangement, i.e., for small $|\qv|$, and before using
it for short wavelengths we have numerically checked its accuracy for the
case of the \ACX\ compounds.

First, self-consistent calculations were performed for two sets of planar
spin spirals with the magnetisation direction inside a Cr sphere fixed by
polar angles $\theta=\pi/2$ and $\phi=\qv\cdot(\tv+\Rv)$, where \tv\ is the
position of a Cr site in the unit cell and \Rv\ is a lattice vector. The
wave vectors \qv=(0,0,$q$) and \qv=($q$,$q$,0) of the spirals varied along
the $\Gamma$--$X$ and $\Gamma$--$K$ high symmetry directions, respectively,
in the range $0\le q\le 2$ in $2\pi/a_0$ units. Obviously, the (0,0,0)
spiral is the collinear FM structure with the magnetisation directed along
the $x$ axis, whereas (2,2,0) and (0,0,2) spirals are equivalent to the
collinear AFZ spin structure with zero net magnetisation discussed in Sec.\
\ref{sec:lsda}. The magnetisation directions in other atomic spheres
were determined self-consistently. 

\begin{figure}[tbp!]
\begin{center}
\includegraphics[width=0.44\textwidth]{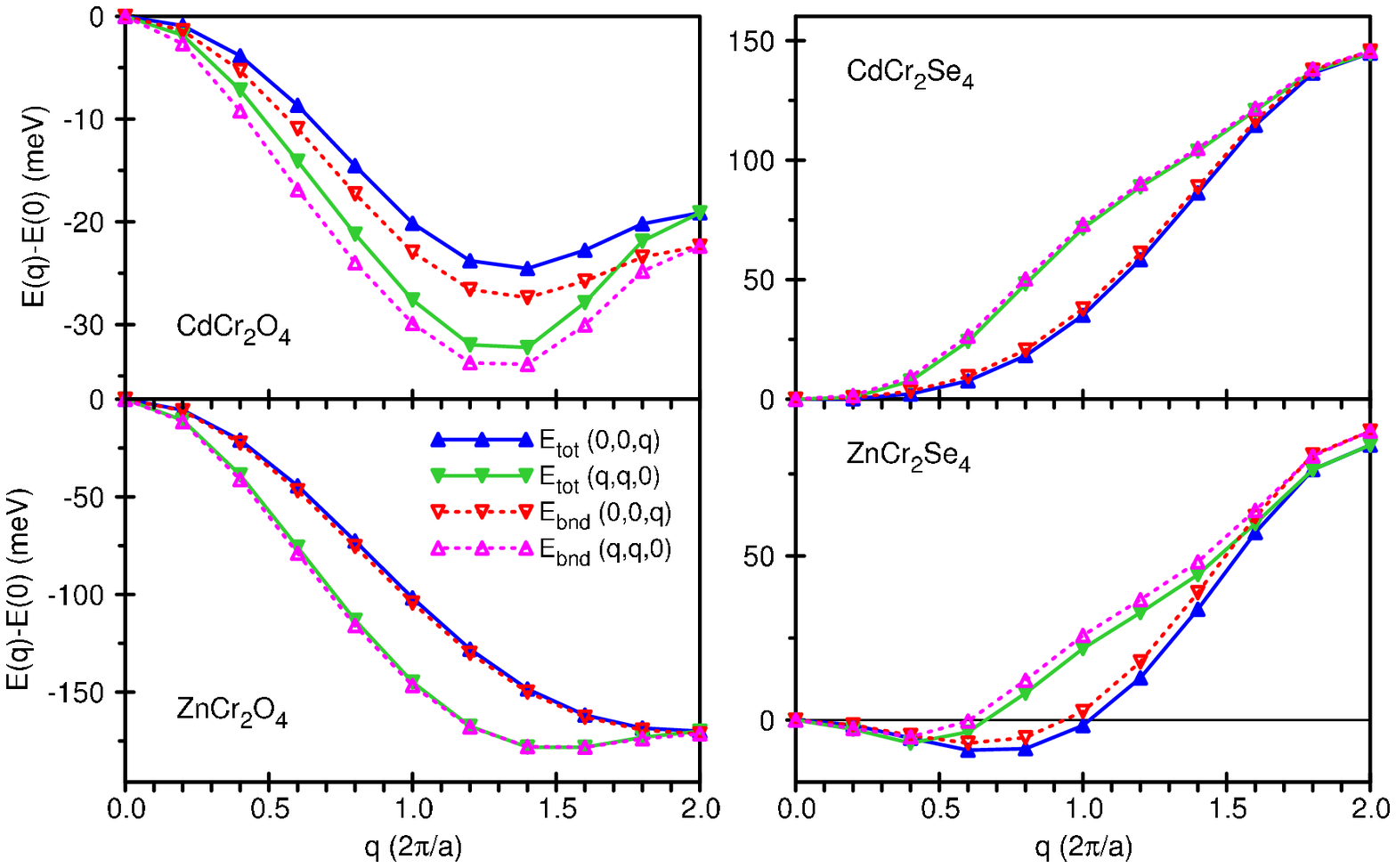}
\end{center}
\caption{\label{fig:Eft2t}(Color online) The dependence of the total \Etq\ 
and band \Ebq\ energies calculated within the LSDA on the wave vector of
(0,0,$q$) and ($q$,$q$,0) spin spirals. The energies are plotted relative to
the energy of the FM states with $q$=0.}
\end{figure}

Then, band energies for the same spin spirals were calculated starting from
self-consistent FM electron densities for \ACSS\ and from the densities for
the ZM solution in the case of oxides. When using LFT for spinels one needs
to specify the magnetisation direction not only for magnetic Cr ions but
also in $A$, $X$, and $E$ spheres in which small but finite magnetic
moments are induced. In a general case this direction is not uniquely
determined by \qv\ but instead depends in a nontrivial way on the
orientation of Cr magnetic moments. We found that good agreement between
the \qv-dependencies of the total \Etq\ and band \Ebq\ energies can be
obtained when the small exchange splitting in $A$, $X$, and $E$ spheres is
completely neglected. \Ebq\ curves calculated in this way for some of the
Cr spinels are compared to \Etq\ in Fig.\ \ref{fig:Eft2t}. Both energies
are plotted relative to the energy of the FM solution with $q$=0. The
difference between \Ebq\ and \Etq\ does not exceed 5 meV and is much
smaller than the variation of the total energy with $q$ even for
\CCO. Somewhat larger differences of about 8 meV are calculated for
sulfides (not shown in Fig.\ \ref{fig:Eft2t}). If the spin-polarisation
inside non-Cr spheres was not switched-off and the magnetisation directions
obtained from self-consistent calculations were used in LFT calculations,
slightly larger differences between the total and band energies were
obtained. Below, when discussing exchange coupling constants between Cr
spins, we will drop the subscript $bnd$ and denote the band energy
calculated with the spin-polarisation in $A$, $X$, and $E$ spheres set to
zero simply as \Eq.

The calculations for spin spirals confirm the conclusion, drawn in Sec.\
\ref{sec:lsda} on the base of comparison of the FM and ZM total energies,
that the energy of the zero net moment solution ($q$=2) is lower than the
FM one ($q$=0) in oxides whereas in \ACSS\ the FM solution becomes more
stable. However, the shift of the minimum of \Eq\ curves to intermediate
values of $q$, most clearly seen for the $\Gamma$--$K$ direction, witnesses
that alongside geometrical frustrations there is also a competition between
the nearest-neighbor exchange interaction and more distant ones. Only in
\CCSe\ and \HCSe\ the minimum is at $q$=0 and the FM solution has the
lowest energy.

\begin{figure}[tbp!]
\begin{center}
\includegraphics[width=0.3\textwidth]{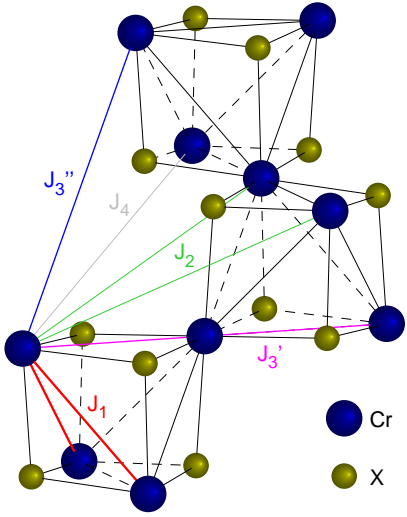}
\end{center}
\caption{\label{fig:Jnn}(Color online) Exchange coupling constants $J_n$
up to the 4-th Cr neighbors.}
\end{figure}

Quantitative estimates for $J_n$ up to the 4-th shell of Cr neighbors were
obtained by mapping the energy of spin spirals onto a classical Heisenberg
model
\begin{equation}
E_{\text{H}}=\frac{1}{4}\sum_{i=1}^{4}\sum_{n=1}^{4}\sum_{j=1}^{z_n}
J_n \Sv_i \cdot \Sv_{j}\,,
\label{eq:emodel}
\end{equation}
where $i$ numbers Cr sites in the unit cell and $j$ runs over $z_n$
neighbors in the $n$-th shell around the site $i$. Positive $J_n$
correspond to AFM coupling between Cr spins. Note that an additional factor
of 1/2 appears in Eq.\ (\ref{eq:emodel}) because the magnetic energy is
given per formula unit and there are two formula units in the unit
cell. The length of the Cr spins $\Sv_i$ was fixed to 3/2. Then, \EHq\
depends only on the angle between a pair of Cr spins which is uniquely
determined by the wave vector \qv\ of a spiral provided that for 4 Cr ions
at positions $\tv_i$ in the unit cell the phases are fixed by
$\phi_i=\tv_i\cdot\qv$.

Some pairs of Cr ions coupled by $J_1$--$J_4$ are shown in Fig.\
\ref{fig:Jnn}.  Each Cr site has 6 first ($J_1$) and 12 second ($J_2$)
neighbors at distances $a_0\sqrt{2}/4$ and $a_0\sqrt{6}/4$,
respectively. The 3-rd shell consists of 12 Cr sites at $a_0\sqrt{2}/2$
which are split into two inequivalent sextets. The corresponding coupling
constants are denoted as $J'_3$ and $J''_3$. $J'_3$ couples Cr sites which
lie on one of $\langle 110 \rangle$ chains and are actually the 2-nd Cr
neighbors along the chain. Cr sites coupled by $J''_3$ belong to parallel
Cr chains. For the spin spirals considered here terms proportional to
$J'_3$ and $J''_3$ in Eq.\ (\ref{eq:emodel}) have the same \qv-dependence
so that it was not possible to separate their contributions to the magnetic
energy and only their average $J_3=(J'_3+J''_3)/2$ could be determined from
the fit. We will return to the discussion of $J'_3$ and $J''_3$
later. Finally, in the 4-th shell there are 12 Cr sites at the distance
$a_0\sqrt{10}/4$ ($J_4$) which lie at the same chains as the 2-nd Cr
neighbors.

\begin{figure}[tbp!]
\begin{center}
\includegraphics[width=0.45\textwidth]{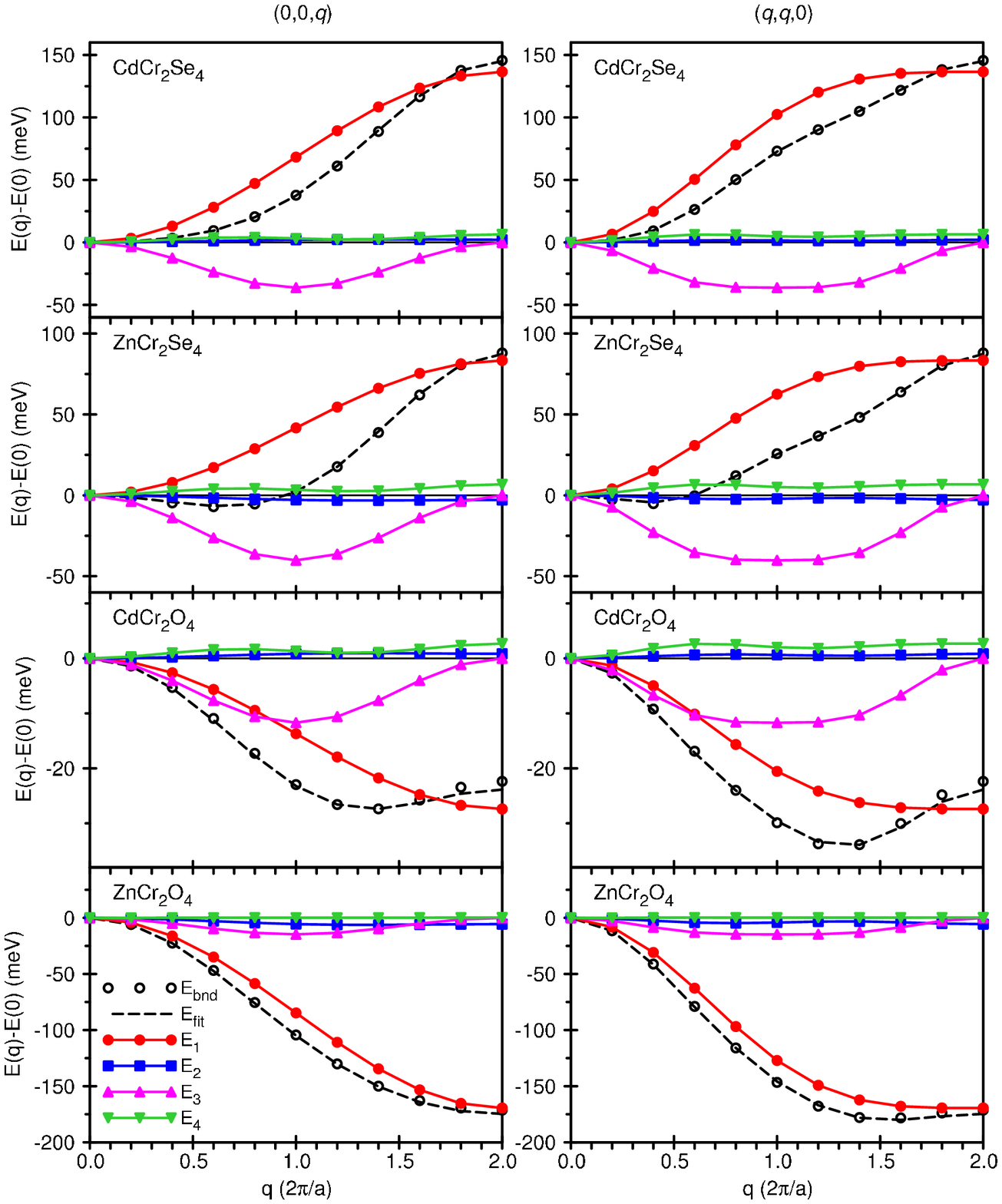}
\end{center}
\caption{\label{fig:Jnn_fit}(Color online) The comparison of calculated
band energies ($E_{\text{bnd}}$) and the results of the fit using Eq.\
(\ref{eq:emodel}) ($E_{\text{fit}}$) for \qv=(0,0,$q$) (left panels) and
\qv=($q$,$q$,0) (right panels). Individual contributions $E_n$ of the terms
proportional to $J_n$ are also plotted.}
\end{figure}

In order to get reliable values of $J_n$, additional LFT calculations for
spirals with wave vectors ($q$,$q$,$q$), (1,$q$,0), and (1,1,$q$) were
performed.  For the (1,$q$,0) and (1,1,$q$) spirals the contributions to
$E_{\text{H}}(\qv)$ proportional to $J_3$ and $J_1$, respectively, do not
depend on $q$, which allows to determine these coupling constants with
higher accuracy. Explicit expressions for $E_{\text{H}}(q)$ for the above
mentioned spin spirals are given in Appendix.

The results of a simultaneous least-squares fit of \EHq\ given by Eq.\
(\ref{eq:emodel}) to \Eq\ calculated within LSDA along the five \qv\
directions are shown in Fig.\ \ref{fig:Jnn_fit} together with contributions
to \EHq\ coming from different nearest neighbors shells ($E_n$).  The
values of $J_{n}$ obtained from the fit are collected in Table
\ref{tab:Jnn_lsda}. The dominant contribution to \EHq\ is provided by the
nearest neighbor coupling $J_1$ which is AFM in \ACO\ and FM in \ACSS. In
oxides, however, the strength of AFM $J_1$ decreases dramatically with the
increase of the $A$ ionic radius: in \CCO\ $J_1$ is more than 5 times
weaker than in \ZCO, while in \HCO\ it becomes almost zero. In \ACSS\ the
strength of FM $J_1$ tends to increase in the row
Zn$\rightarrow$Cd$\rightarrow$Hg but the changes are not as strong as in
the oxides; with the values of $J_1$ calculated for the Cd and Hg compounds
being comparable.

Another significant contribution to \EHq\ comes from the term proportional
to $J_3$ which is AFM in all the \ACX\ spinels considered in the present
work. It is the competition between the $J_1$ and $J_3$ terms which shifts
the minimum of the (0,0,$q$) and ($q$,$q$,0) curves to incommensurate \qv\
vectors. For \qv=(0,0,$q$) \EHq\ has the same $q$-dependence as a linear
chain of classical spins with competing nearest $J=J_1$ and next-nearest
$J'=2J_3$ neighbor interactions. The ground state of such a chain becomes
incommensurate if the ratio $J'/J$ is larger than the critical value of
0.25. Since $J_3$ is not very sensitive to the size of $A$ ion, the ratio
$2J_3/J_1$, which is less than 0.1 in \ZCO, increases to 0.45 in \CCO,
while in the Hg oxide $J_3$ becomes the dominant magnetic interaction. The
values of $J_3$ calulated for \ACSS\ are more than 2 times larger than in
the oxides. The largest $|2J_3/J_1|$ ratios of 0.75 and 0.48 are found for
\ZCS\ and \ZCSe, in which the FM nearest neighbor interaction is the
weakest. In other two selenides $|2J_3/J_1|\approx 0.26$ is only slighly
larger than the critical value of 0.25.

According to the results of the the least-square fit the exchange
interactions between the 2-nd and 4-th Cr neighbors are much weaker than
$J_1$ and $J_3$. While the sign of $J_2$ varies form one compound to
another, $J_4$ is always ferromagnetic and somewhat larger in sulfides and
selenides compared to oxides.

The reliability of the fitted $J_n$ was checked by comparing the values
obtained from the fit to the band energies calculated for all 5 spin
spirals ($J_n^{(5)}$) to those fitted to (0,0,$q$) and ($q$,$q$,0) results
only ($J_n^{(2)}$). In the case of $J_1$, the relative uncertainty of the
determination of $J_n$, defined as
$\delta_n=|J_n^{(5)}-J_n^{(2)}|/J_n^{(5)}$, is less than 1\%, except for
\HCO\ where $J_1$ itself is very small. The uncertainty in $J_3$ values is
about 1\% in \ACSS\ and 10\% in oxides. Tha values of $\delta_2\sim5$\% and
$\delta_4\sim10$\% are also smaller in \ACSS\ than in the oxides, in which
they are about 30\%. The largest uncertainties in the LSDA values of
$J_2$--$J_4$ were found in the case of \ZCO. A plausible reason is that the
nearest-neighbors terms of the order of $(\Sv_i \cdot \Sv_j)^2$ in the
expansion of the magnetic energy \cite{RJ97} may become comparable to the
contribution of weak couplings between distant Cr neighbors. Neither the
fitted values of $J_n$ nor the quality of the fit were noticeably affected
if exchange couplings between 5-th Cr neighbors were included into the
fit. Accounting for more distant 6-th neighbors allowed to decrease the
mean deviation between the calculated and fitted energies for \ACSS\ but
the obtained values of $J_6$ were even smaller than $J_2$ and $J_4$ and
could not be determined reliably.  Finally, significantly smaller values of
$\delta_n$ were obtained when fitting \EHq\ to LSDA+$U$ band energies.

\begin{table}
\caption{\label{tab:Jnn_lsda}Exchange coupling constants $J_n/\kb$ (K) and
\TCW\ (K) obtained from the least-squares fit of \EHq\ given by Eq.\
(\ref{eq:emodel}) to the energy of spin spirals calculated within LSDA. }
\begin{ruledtabular}
\begin{tabular}{crrrrr}
   & \multicolumn{1}{c}{$J_1/\kb$}
  &  \multicolumn{1}{c}{$J_2/\kb$} & \multicolumn{1}{c}{$J_3/\kb$}
  &  \multicolumn{1}{c}{$J_4/\kb$} & \multicolumn{1}{c}{\TCW} \\
\hline
\ZCO & 109 &  1.8 &  4.8 & -0.1 & -916 \\
\CCO &  18 & -0.3 &  3.8 & -0.9 & -172 \\
\HCO &  -1 &  1.8 &  5.4 & -0.8 &  -92 \\
\hline				      	        
\ZCS & -37 &  2.8 & 13.8 & -1.5 &   49 \\
\CCS & -74 &  0.6 & 12.0 & -1.7 &  392 \\
\HCS & -86 &  2.8 & 13.3 & -1.7 &  432 \\
\hline				      	        
\ZCSe& -54 &  0.9 & 13.0 & -2.2 &  228 \\
\CCSe& -88 & -0.7 & 11.7 & -2.1 &  526 \\
\HCSe& -86 &  0.0 & 11.6 & -2.6 &  509 \\
\end{tabular}
\end{ruledtabular}
\end{table}

$J_n$ obtained by fitting \EHq\ to band energies calculated using the
LSDA+$U$ approach with $U$=2, 3, and 4 eV are presented in Table
\ref{tab:Jnn_U}. Both $J_1$ and $J_3$ derived from the \LDAU\ calculations
with $U$=2 eV for \ACO\ are weaker than the corresponding LSDA values. The
only exception is \HCO, in which $J_1$ is weakly FM in LSDA and becomes
somewhat stronger in \LDAU. In \ZCO\ the value of $J_1$ rapidly decreases
with the increase of $U$ as it is expected if $J\sim t^2/U$. In Cd and Hg
spinels, on the other hand, the increase of $U$ tends to make $J_1$ more
ferromagnetic. It looks as if there were two competing contributions of the
opposite signs to $J_1$: as the AFM one is suppressed by $U$ the
ferromagnetic contribution wins. Comparable values of $J_1$=0.5 meV,
$J_2\approx$0 meV, and $J_3=0.15$ meV were obtained from \LDAU\
calculations for \CCO\ in Ref.\ \onlinecite{CFT06}. These calculations,
however, were performed using theoretical lattice parameters.

In \ZCS\ $|J_1|$ calculated with $U$=2 eV is slightly larger than the LSDA
value and continue to increase with the increase of $U$, whereas in \ZCSe\
$J_1$ is practically independent of $U$. In other \ACSS\ spinels switching
on $U$ suppresses $J_1$ but the increase of $U$ from 2 to 4 eV has only
minor effect on its strength.

In all the compounds $J_3$ obtained from the \LDAU\ calculations is AFM and
weaker than in LSDA. In \ACO\ and \ACS\ its value decreases with the
increase of $U$ while in selenides its $U$ dependence is very weak. $J_2$
remains vanishingly small but shows clear tendency to become more FM as $U$
increases. Finally, in \ACO\ and \ACSe\ the increase of $U$ affects $J_4$
in opposite ways: in oxides the strength of FM $J_4$ increases whereas in
selenides $|J_4|$ decreases and for $U$=4 eV it changes sign.

\begin{table}
\caption{\label{tab:Jnn_U}Exchange coupling constants $J_n/\kb$ (K) and
\TCW\ (K) obtained from the least-squares fit of \EHq\ given by Eq.\
(\ref{eq:emodel}) to the energy of spin spirals calculated using the \LDAU\
approach with $U$=2, 3, and 4 eV.}

\begin{ruledtabular}
\begin{tabular}{ccrrrrr}
\ACX  & $U$ & \multicolumn{1}{c}{$J_1/\kb$}
  &  \multicolumn{1}{c}{$J_2/\kb$} & \multicolumn{1}{c}{$J_3/\kb$}
  &  \multicolumn{1}{c}{$J_4/\kb$} & \multicolumn{1}{c}{\TCW} \\
\hline
\ZCO &2 eV&  61 &  0.3 &  2.9 & -0.3 & -500 \\
     &3 eV&  40 &  0.0 &  2.3 & -0.4 & -328 \\
     &4 eV&  25 & -0.2 &  1.9 & -0.5 & -209 \\
\CCO &2 eV&   6 & -0.3 &  2.2 & -0.4 &  -64 \\
     &3 eV&  -4 & -0.4 &  1.7 & -0.4 &   12 \\
     &4 eV&  -9 & -0.4 &  1.4 & -0.4 &   62 \\
\HCO &2 eV&  -7 &  1.2 &  3.3 &  0.0 &  -14 \\
     &3 eV& -14 &  0.6 &  2.4 & -0.1 &   59 \\
     &4 eV& -18 &  0.3 &  1.9 & -0.2 &  104 \\
\hline				           	     
\ZCS &2 eV& -43 &  1.7 &  8.5 & -0.6 &  175 \\
     &3 eV& -48 &  0.8 &  7.2 & -0.8 &  267 \\
     &4 eV& -52 &  0.1 &  6.4 & -0.7 &  306 \\
\CCS &2 eV& -62 &  0.4 &  7.2 & -0.9 &  367 \\
     &3 eV& -65 & -0.2 &  6.2 & -0.9 &  416 \\
     &4 eV& -66 & -0.6 &  5.7 & -0.7 &  433 \\
\HCS &2 eV& -72 &  1.6 &  8.0 & -0.9 &  412 \\
     &3 eV& -74 &  0.5 &  6.6 & -0.9 &  464 \\
     &4 eV& -74 & -0.2 &  5.8 & -0.7 &  481 \\
\hline				           	     
\ZCSe&2 eV& -49 &  0.5 &  8.3 & -0.9 &  246 \\
     &3 eV& -52 & -0.5 &  7.8 & -0.6 &  286 \\
     &4 eV& -52 & -1.1 &  8.1 &  0.0 &  286 \\
\CCSe&2 eV& -70 & -0.5 &  7.1 & -0.9 &  432 \\
     &3 eV& -69 & -1.1 &  7.1 & -0.4 &  434 \\
     &4 eV& -67 & -1.4 &  7.6 &  0.2 &  410 \\
\HCSe&2 eV& -69 & -0.1 &  7.5 & -0.9 &  422 \\
     &3 eV& -68 & -1.0 &  7.2 & -0.2 &  420 \\
     &4 eV& -66 & -1.4 &  7.8 &  1.1 &  384 \\
\end{tabular}
\end{ruledtabular}
\end{table}

In order to separate $J'_3$ and $J''_3$ contributions to \EHq\ we performed
calculations for non-collinear spin superstructures in which Cr moments
only in every second (001) plane, the one that contains [110] Cr chains,
are oriented in the $ab$ plane; with their directions being determined as
before by $\theta=\pi/2$ and $\phi=\qv\cdot(\tv+\Rv)$. Cr moments in other
planes, which contain [1$\bar{1}$0] chains, are aligned parallel to the $c$
axis ($\theta=0$) so that their directions do not depend on $\phi$ and,
consequently, on \qv. In this case the 3-rd Cr neighbors in the $ab$ plane
that lie on the same [110] chain ($J'_3$) and on parallel chains ($J''_3$)
give different contributions to the $q$-dependence of the energy of
($q$,$q$,0) and ($q$,$-q$,0) spin spirals, which allows to determine $J'_3$
and $J''_3$ separately provided that other $J_n$ are known. Calculations
performed for \ACO\ and \ACSe\ show that $J'_3$ and $J''_3$ are of
comparable strengths. The calculated $J''_3/J'_3$ ratio is 0.6, 0.5, and
1.1 in Zn, Cd, and Hg oxides, respectively, while for the selenides the
values of 0.8, 0.8, and 1.0 were obtained. It seems plausible that the
contribution of Hg 6$s$ states to Cr $d$--$X$ $p$--Cr $d$ hybridisation is
responsible for somewhat larger values of $J''_3$ in the Hg compounds.

It should be mentioned that the values of $J_n$ presented in Table
\ref{tab:Jnn_U} are somewhat different from the preliminary results
published in Ref.\ \onlinecite{YA07} for the following reasons: i) for
\CCO\ old structural data from Ref.\ \onlinecite{VH47} with the O
fractional coordinate $x$=0.260, which seems to be too small, were used;
ii) the calculations for oxides in Ref.\ \onlinecite{YA07} were performed
with O $s$ and $p$ states only included in the LMTO basis set. This gives
smaller values of $J_1$ compared to the present calculations in which O $d$
are also included into the basis; iii) the calculations in Ref.\
\onlinecite{YA07} were performed for ($q$,$q$,0) and (0,0,$q$) spirals only
and the coupling constants between 3-rd Cr neighbors along the $\langle 110
\rangle$ chains were included into the fit instead of $J_4$. Later it was
verified that accounting for true $J_4$ to 4-th Cr neighbors improves the
quality of the fit and seems more physical. We have to stress, however,
that despite the differences in calculated values of $J_n$ the main
conclusions of Ref.\ \onlinecite{YA07} concerning the relative strengths of
various exchange interactions and their origins remain unaffected.

\section{\label{sec:discus}Discussion}

The effective Curie-Weiss temperatures estimated from the calculatated
$J_n$ according to
\begin{equation}
\TCW=\frac{S(S+1)}{3}\sum_n z_n J_n\,,
\label{eq:TCW}
\end{equation}
with $S$=3/2 are given in the last column of Tables \ref{tab:Jnn_lsda} and
\ref{tab:Jnn_U}. The comparison of the estimated \TCW\ to experimental
values from Ref.\ \onlinecite{RKMH+07} (Table \ref{tab:str}) shows that the
LSDA calculations reproduce the opposite signs of \TCW\ in \ACO\ and \ACSS\
but strongly overestimate its magnitude, which is not surprising as one can
expect that LSDA underestimates correlation effects in the rather localised
Cr 3$d$ shell. Since the absolute values of \TCW\ calculated for oxides
rapidly decrease with the increase of $U$ in the \LDAU\ calculations, it is
possible to adjust the value of $U$ so as to reproduce the experimental
\TCW. This, however, would require to use different values of $U$ for the
Zn, Cd, and Hg oxospinels. In \ACSS\ the estimated values of \TCW, which
are too high already in LSDA, further increase when the \LDAU\ approach is
used. As the dominant contribution to \TCW\ is provided by $J_1$, it seems
that the tendency to ferromagnetic coupling between nearest Cr neighbors
is overestimated in the \LDAU\ calculations.

The comparison of the calculated $J_n$ to experimental estimates for the
exchange coupling constants in \ZCS\ ($J_1$=2.66 K, $J_2$=$-$1.15 K,
$J_3$=0.29 K, $J_4$=0.13 K) \cite{HHSC95} and \CCS\ ($J_1$=13.25 K,
$J_3$=$-$0.915 K) \cite{PAFN92} also shows that the calculations
overestimate the strength of the exchange interactions between Cr
spins. Note that the experimental values of $J_n$ should be multiplied by a
factor of 2 because of the different definition of the magnetic energy [see
Eq.~(2) in Ref.\ \onlinecite{HHSC95}], while theirs signs should be
inverted since in Refs.\ \onlinecite{HHSC95,PAFN92} the positive sign of
$J_n$ corresponds to FM coupling. Also it is worth mentioning that $J_n$
are not directly measurable quantities and in order to determine them from
experimental data additional assumption should be made. The experimental
wave vector $\qv_{exp}=$(0,0,0.79) of the helimagnetic structure in \ZCS\
is very close to $\qv_{min}=$(0,0,0.82) at which the LSDA energy of the
(0,0,$q$) spin spiral has a minimum. Under assumption that $J_2$ and $J_4$
are much smaller than $J_1$ and $J_3$ this gives the $J_1/J_3$ ratio of
0.39 and 0.37 in the experiment and theory, respectively. For \ZCSe\ LSDA
gives the minimum of $E(q)$ along the $\Gamma$--$X$ at
$\qv_{min}=$(0,0,0.65) which is larger than the experimental wave vector
$\qv_{exp}=$(0,0,0.466). \cite{ASSI+78} In the \LDAU\ calculations,
however, the minimum shifts towards smaller $q$ values.
 
In order to understand better the dependence of the calculated exchange
coupling constants on the chemical composition and on the value of the
parameter $U$ in \LDAU\ calculations, let us for simplicity consider a pair
of Cr$^{3+}$ ions with occupied \tgu\ states at zero energy, unoccupied
\tgd\ states at the energy \dtud, and an effective \tg--\tg\ hopping matrix
element $t$. The splitting $\dtud$ is close to $3\JH$ in the LSDA and
$3J_{\text{H}}+U'$ in the \LDAU\ calculations. Cr \egs\ states with both
spin projections are unoccupied and also split by \deud\ which in LSDA is
equal to 3\JH. The LSDA energy difference \degt\ between \egs\ and \tg\
states with the same spin projections is equal to the crystal field
splitting \dcf\ which is of the same order as \deud\ as is evidenced by the
results presented in Sec.\ \ref{sec:lsda}.

\begin{figure}[tbp!]
\begin{center}
\includegraphics[width=0.44\textwidth]{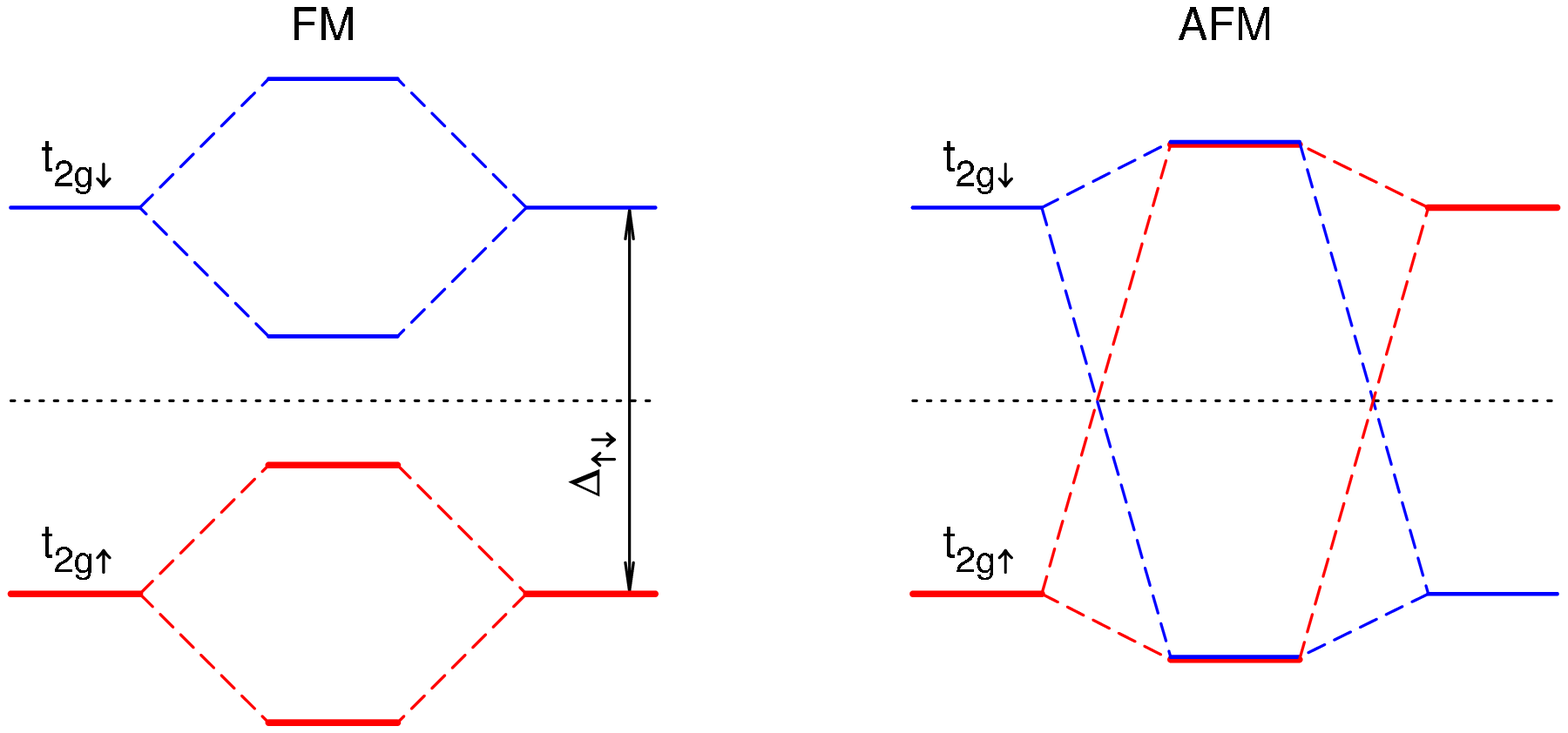}
\end{center}
\caption{\label{fig:t2gt2g}(Color online) A sketch of bonding and
antibonding states between Cr \tg\ state for FM (left) and AFM (right)
alignments of Cr magnetisations. The Fermi level is plotted by a horizontal
dotted line.}
\end{figure}

Hybridisation, either direct or indirect via $X$ $p$ states, between the
half-filled Cr \tg\ states results in AFM exchange coupling provided that
the \tg\ bandwidth is less than \dtud. Indeed, if Cr moments are aligned
ferromagnetically, the hybridisation between the \tg\ states gives no gain
in the kinetic energy as in the majority-spin chanel both bonding and
antibonding combinations with energies $\ebu=-t$ and $\eau=t$ are occupied,
whereas in the minority-spin chanel both of them
($\varepsilon^{b,ab}_{\downarrow}=\dtud\pm t$) are empty (see Fig.\
\ref{fig:t2gt2g}). If the magnetisations of two Cr ions are antiparallel,
an occupied \tgu\ state can hybridise only with the unoccupied \tg\ state
with the same spin projection of another ion lying at the energy \dtud. In
this case only their bonding combination with the energy $\ebu\approx
-t^2/\dtud$ is occupied. Since the energy of the bonding state with the
opposite spin projection is exactly the same, this gives
\begin{equation}
E_{FM}-E_{AFM}\equiv 2J_{AFM}S^2\approx 2t^2/\dtud
\label{eq:jafm}
\end{equation}
for the energy difference of the FM and AFM configurations.

\begin{figure}[tbp!]
\begin{center}
\includegraphics[width=0.44\textwidth]{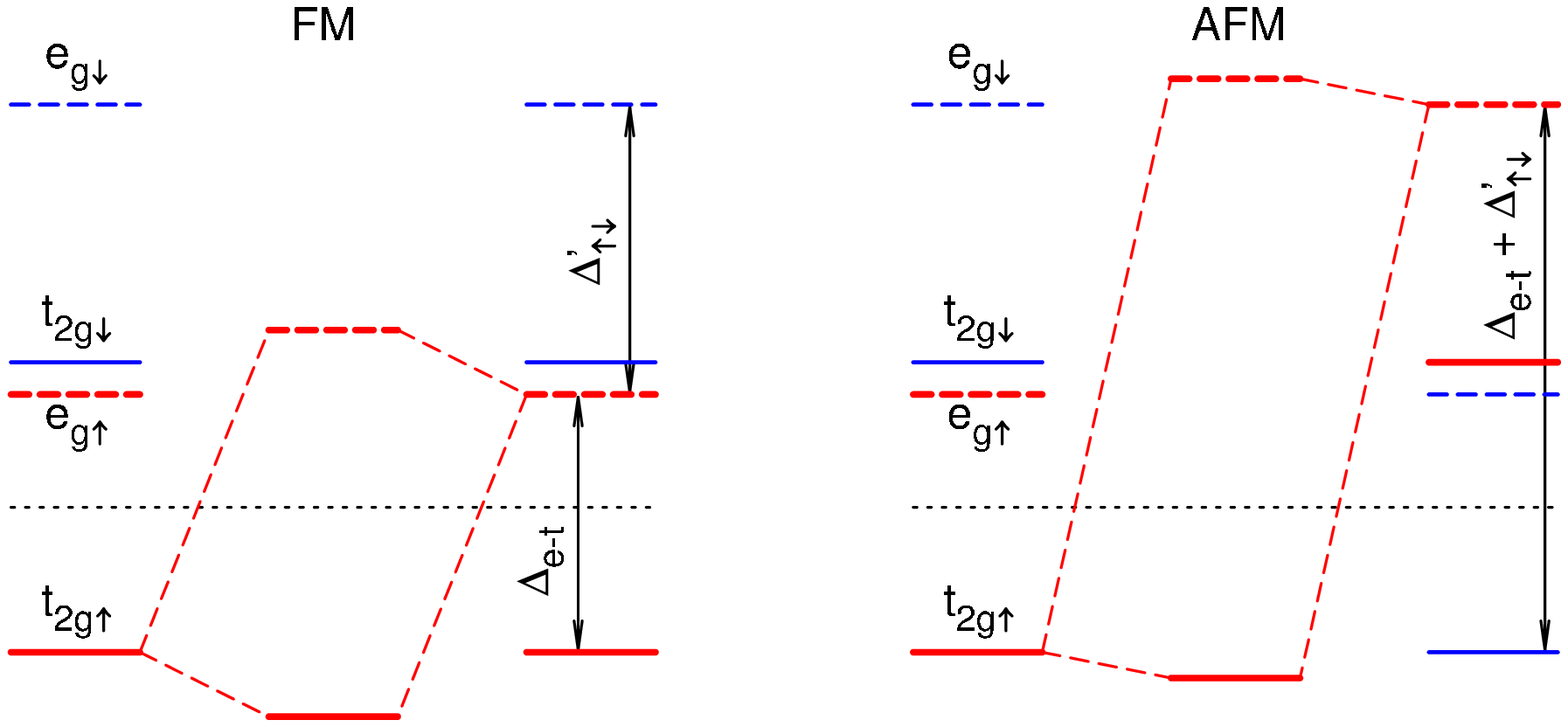}
\end{center}
\caption{\label{fig:t2geg}(Color online) A sketch of bonding and
antibonding states between an occupied Cr \tgu\ state and an unoccupied
\egs\ state with the same spin projection for FM (left) and AFM (right)
alignments of Cr magnetisations. The Fermi level is plotted by a horizontal
dotted line.}
\end{figure}

Cr \eg\ states with both spin projections are unoccupied and \tg--\egs\
hybridisation ($t'$) lowers the energy of the FM as well as AFM
configuration because in both cases only the bonding combination is
occupied (see Fig.\ \ref{fig:t2geg}). The energy of the bonding state
$\ebu\approx -t'^{2}/(\varepsilon_{e_{g}}-\varepsilon_{t_{2g}})$ is,
however, lower if the magnetisations of two Cr ions are parallel. In this
case the \egs\ partner with the same spin as the occupied \tg\ state has
lower energy $\varepsilon_{e_{g}}=\degt$ vs.\
$\varepsilon_{e_{g}}=\degt+\deud$ in the AFM case, so that
\begin{equation}
E_{FM}-E_{AFM}\equiv2J_{FM}S^2
\approx -\frac{2t'^{2}\deud}{\degt(\degt+\deud)} \,,
\label{eq:jfm}
\end{equation}
and the effective coupling $J_{FM}$ is ferromagnetic. It should be
mentioned that these estimates for the effective exchange couplings are in
agreement with the well known Goodenough-Kanamori rules. \cite{DM68}

Thus, the strength and even the sign of $J_1$ in \ACX\ are determined by
the competition of two contributions of opposite signs, $J_{AFM}$ and
$J_{FM}$, which depend on the strength of the \tg--\tg\ and \tg--\eg\
hybridisation, respectively. In \ZCO\ $J_{AFM}$ dominates because of the
strong direct hopping of $dd\sigma$ type between the Cr \tg\ states. This
conclusion has been confirmed numerically by the results of test LSDA
calculations in which the hybridisation between Cr $d$ states was switched
off. $J_1$ derived from a fit to band energies calculated in this way is
about 1 K, whereas for other $J_n$ the values comparable to those given in
Table \ref{tab:Jnn_lsda} were obtained. With the increase of the lattice
constant in the row Zn$\rightarrow$Cd$\rightarrow$Hg the strength of the
direct $d$--$d$ hopping and, as a consequence, of $J_{AFM}$ rapidly
decreases so that in \HCO\ according to the LSDA results $J_{AFM}$ and
$J_{FM}$ almost exactly compensate each other. The dependence of $J_1$ on
the value of the parameter $U$ in the \LDAU\ calculations can be explained
by weakening of $J_{AFM}$ caused by the increase of the denominator in Eq.\
(\ref{eq:jafm}) which in this case is given by $\dtud\approx 2\JH+U$.

In \ACSS\ spinels the distance between Cr nearest neighbors (see Table
\ref{tab:str}) is much larger than in \ACO\ and the contribution of the
direct \tg--\tg\ hopping to $J_1$ becomes less significant. For \ZCSe, for
instance, calculations with switched-off Cr $d$--$d$ hybridisation give the
value of FM $J_1$ which is only 5\% larger than the LSDA one.  Thus, the
indirect hybridisation via $X$ $p$ states between the half-filled \tg\ and
empty \eg\ states split by the on-site exchange interaction provides the
dominant contribution to FM $J_1$ as was proposed by J.~Goodenough in Ref.\
\onlinecite{Goodenough69}. A direct numerical proof is provided by test
calculations in which an external orbital-dependent potential $V_{eg}$ was
added to Cr \egsu\ states; with its strength $V_{eg}\approx3J_{\text{H}}$
being equal to the LSDA exchange splitting of the \egs\ states. Then, the
energies of Cr \egs\ states with both spin projections are equal so that
$\deud\approx0$ and the nominator in the expression (\ref{eq:jfm}) for
$J_{FM}$ becomes zero.  $J_1$=15 K estimated from such calculations for
\ZCSe\ changes its sign to AFM and its absolute value is significantly
smaller than $|J_1|$=54 K obtained from the LSDA calculations. This
indicates that the indirect \tg--$X$ $p$--\tg\ hybridisation also
contribute to $J_1$ but it is much weaker than $J_{FM}$. Similar
calculations for \ZCO\ result in substantial increase of the strength of
AFM $J_1$ which means the FM contribution is present also in oxides but
$J_{AFM}$ wins due to direct $d$--$d$ hybridisation.

In order to explain the weird behavior of $J_1$ in \LDAU\ calculations one
needs to analyze in more details various parameters entering Eq.\
(\ref{eq:jfm}). Recalling that the orbital dependent part of the \LDAU\
potential can be approximated as $V_{i}=U'(1/2-n_i)$ with $U'=U-\JH$ one
gets
\begin{eqnarray}
\deud&=& 3\JH+U'(n_{\uparrow}-n_{\downarrow}) \label{eq:dexu} \,, \\
\degt&=&\dcf+U'(1-n_{\uparrow})  \label{eq:dcfu} \,,
\end{eqnarray}
where $n_{\uparrow}$ and $n_{\downarrow}$ are the occupation numbers of
\egsu\ and \egsd\ states, respectively. Here we neglected the small
($\sim$0.05) deviation of the occupation numbers of the minority-spin \tg\
states from unity. Then, Eq.\ (\ref{eq:jfm}) can be written as
\begin{eqnarray}
2J_{FM}S^2 &\approx& -\frac{2t'^{2}[3\JH+U'(n_{\uparrow}-n_{\downarrow})]}
{\dcf+U'(1-n_{\uparrow})} \nn \\
&\times&
\frac{1}{\dcf+U'(1-n_{\downarrow})+3\JH} \,.
\label{eq:jfmU}
\end{eqnarray}
If the occupations of the Cr \egs\ were equal to zero Eq.\ (\ref{eq:jfmU})
would become
\begin{equation}
2J_{FM}S^2 \approx -\frac{6t'^{2}\JH}{(\dcf+U')(\dcf+U'+3\JH)} \,,
\label{eq:jfmUion}
\end{equation}
which would lead to fast decrease of $J_{FM}$ with the increase of $U'$.
However, since $n_{\uparrow}\sim 0.45$ and $n_{\downarrow}\sim 0.25$ were
calculated for \ACSS\ the appearance of the term proportional to
$U'(n_{\uparrow}-n_{\downarrow})$ in the nominator of Eq.\ (\ref{eq:jfmU})
slows down the decrease of $J_{FM}$. Also the terms proportional to
$n_{\uparrow}$ and $n_{\downarrow}$ in the dominator of Eq.\
(\ref{eq:jfmUion}) effectively decrease $U'$, thus leading to the increase
of $J_{FM}$ as compared to purely ionic model with zero occupations of the
\egs\ states. This is one of possible reasons why the FM contribution to
$J_1$ in overestimated in the \LDAU\ calculations.

The calculated behavior of $J_2$, $J_3$, and $J_4$ follows the predictions
made by K.~Dwight and N.~Menyuk in Refs.\ \onlinecite{DM67,DM68} on the
base of the Goodenough-Kanamori rules of
superexchange.\cite{Goodenough58,Kanamori59} Hybridisation paths Cr
$d$--$X$ $p$--$X$ $p$--Cr $d$ responsible for these interactions include an
intermediate hopping between $p$ states of two $X$ ions. This reduces the
strength of the superexchange interactions but does not change the signs of
\tg--\tg\ and \tg\--\eg\ contributions to $J_n$, which remain antiferro-
and ferromagnetic, respectively. Indirect \tg--\tg\ and \tg--\eg\ hoppings
give comparable contributions to $J_2$. Being of opposite signs, these
contributions compensate each other. This explains the smallness of $J_2$
and the variation of its sign in the \ACX\ spinels. Similar hybridisation
paths between $d$ states of 3-rd neighbors, with the dominant contribution
of the indirect \tg--\tg\ hopping, lead to the appearance of AFM $J'_3$ and
$J''_3$. In the real structure the distance between the pair of $X$ ions
along the $J'_3$ path is smaller than along the $J''_3$ one; the difference
being proportional to the deviation of the positional parameter $x$ from
1/4. This may explain why the calculated values of $J'_3$ are somewhat
larger than $J''_3$. On the other hand, since the $J''_3$ pair of $X$ ions
belongs to the same $AX_4$ tetrahedron, $J''_3$ is more sensitive to $A$
$s$--$X$ $p$ hybridisation, which may be responsible for the increase of
the $J''_3/J'_3$ ratio in the Hg-containing spinels. Finally, indirect
hybridisation between \tg\ and \eg\ states gives dominant contribution to
the superexchange between 4-th neighbors \cite{DM68} which leads to FM
$J_4$.

\section{\label{sec:summ}Summary and conclusions}

The LSDA($+U$) band structure calculations performed for some \ACX\ spinels
reproduce the change of the sign of the nearest-neighbor exchange
interaction between Cr spins from AFM in \ACO\ compounds to FM in \ACSS\
ones. It has been verified that the strength and the sign of $J_1$ depend
on relative strengths of two contributions of opposite signs, AFM and FM,
which are due to Cr \tg--\tg\ and \tg--\eg\ hybridisations,
respectively. In \ZCO\ $J_{AFM}$ dominates because of the strong direct
hopping between the Cr \tg\ states. As Zn is replaced by Cd or Hg, the
strength of the direct hopping rapidly decreases with the increase of the
Cr--Cr separation. As a result, $J_1$ in the corresponding oxides is much
weaker than in \ZCO. In the \ACSS\ spinels the FM contribution originating
from the indirect hopping between half-filled \tg\ and empty \eg\ states,
split by the on-site exchange interaction, becomes dominant. However, the
magnitute of the FM contribution to $J_1$ seems to be overestimated by the
calculations. The strongest among more distant exchange couplings are
$J'_3$ and $J''_3$ to two sextets of 3-rd Cr neighbors. These interactions
are always AFM and of comparable strengths. The couplings to 2-nd and 4-th
Cr neighbors are significantly weaker than $J_1$ and $J_3$. The obtained
results are in accordance with the analysis of the exchange couplings in Cr
spinels based on the Goodenough-Kanamori rules of superexchange and may be
helpful for understanding the magnetic properties not only of Cr spinels
but also Ti (MgTi$_2$O$_4$) and V (ZnV$_2$O$_4$, CdV$_2$O$_4$) ones, in
which orbital degrees of freedom come into
play. \cite{SRRR+04,LLUP+04,NO02}

\begin{acknowledgments}
Author is grateful to P.~Fulde, N.~Shannon, K.~Penc, and V.~Antonov for many
helpful discussions and to H.~Ueda for providing recent experimental data
on the crystal structures of \CCO\ and \HCO\ spinels.
\end{acknowledgments}

\appendix*
\section{\label{sec:app}}{}

Below explicite expressions for the difference
$E_{\text{H}}(\qv)-E_{\text{H}}(0)$, where $E_{\text{H}}(\qv)$ is given by
Eq.\ (\ref{eq:emodel}) and
\begin{equation}
E_{\text{H}}(0)=6J_1+12J_2+6J_3+12J_4\,,
\end{equation}
are written as a functions of the angle $\phi=q\pi/2$ for some directions
of the wave vector \qv\ of spin spirals: \\
\qv=(0,0,$q$) ($\Gamma$--$X$):
\begin{eqnarray}
E(\phi) &=& 4 J_1 (\cos \phi - 1) + 
  4 J_2 (2 \cos \phi + \cos 2\phi - 3)  \nn \\ 
&+& 8 J_3 (\cos 2\phi- 1 ) \nn \\
&+& 4 J_4 (\cos \phi + \cos 3\phi -2) ;
\end{eqnarray}
\qv=($q$,$q$,0) ($\Gamma$--$K$):
\begin{eqnarray}
E(\phi)&=& J_1 (4 \cos \phi + \cos 2\phi -5) \nn \\
&+& 2 J_2 (2\cos\phi + \cos2\phi + 2\cos3\phi - 5) \nn \\
&+& 2 J_3 (4\cos2\phi+ \cos4\phi - 5 ) \nn \\
&+& 2 J_4 (2\cos\phi + \cos2\phi + 2\cos3\phi + \cos 4\phi - 6 ); \nn \\
\end{eqnarray}
\qv=($q$,$q$,$q$) ($\Gamma$--$L$):
\begin{eqnarray}
E(\phi)&=& 3 J_1 (\cos 2\phi - 1) \nn \\
&+& 3 J_2 (2 \cos 2\phi + \cos 4\phi - 3) \nn \\ 
&+& 6 J_3 (\cos 4\phi - 1 ) \nn \\
&+& 6 J_4 (\cos 2\phi + \cos 4\phi - 2) ;
\end{eqnarray}
\qv=(1,$q$,0):
\begin{eqnarray}
E(\phi)&=& 2 J_1 (\cos\phi - 3) - 4 J_2 ( \cos\phi + 3) \nn \\ 
&-& 16 J_3 + 2 J_4 (\cos\phi + \cos3\phi - 6) ;
\end{eqnarray}
\qv=(1,1,$q$):
\begin{eqnarray}
E(\phi)&=& -6J_1-12J_2  \nn \\
&-& 8 J_3 (\cos2\phi+1)-12J_4 \,;
\end{eqnarray}
Here $J_3=(J_3'+J_3'')/2$ is the average of the exchange coupling constants
for two inequivalent groups of 3-rd Cr neighbors.

%\bibliography{./ACr2X4}

\end{document}